\documentclass[twocolappendix]{emulateapj}
\usepackage{longtable, hyperref, graphicx, subfigure,amsmath}
\usepackage{lmodern}
\usepackage[T1]{fontenc}

\newcommand{\Msun}{\ifmmode {M_{\odot}}\else${M_{\odot}}$\fi}
\newcommand{\Rsun}{\ifmmode {R_{\odot}}\else${R_{\odot}}$\fi}
\newcommand{\Lsun}{\ifmmode {L_{\odot}}\else${L_{\odot}}$\fi}
\newcommand{\lapprox }{{\lower0.8ex\hbox{$\buildrel <\over\sim$}}}
\newcommand{\gapprox }{{\lower0.8ex\hbox{$\buildrel >\over\sim$}}}
\def\amin{\ifmmode^{\prime}\else$^{\prime}$\fi}
\def\asec{\ifmmode^{\prime\prime}\else$^{\prime\prime}$\fi}

\def\ROSAT{\it ROSAT}
\newcommand{\degree}{\ifmmode {^\circ}\else$^\circ$\fi}
\newcommand{\halpha}{$H\alpha$\ }
\newcommand{\Ro}{\ifmmode {R_o}\else$R_o\ $\fi}

\slugcomment{Accepted September 18, 2014}
\shorttitle{The Factory \& The Beehive II}
\shortauthors{Douglas et al.}
\begin{document}

\title{The Factory and The Beehive\\II.\ Activity and Rotation in Praesepe and the Hyades} 

\author{S.~T.~Douglas\altaffilmark{1},
M.~A.~Ag{\" u}eros\altaffilmark{1},
K.~R.~Covey\altaffilmark{2},
E.~C.~Bowsher\altaffilmark{1}, 
J.~J.~Bochanski\altaffilmark{3}, 
P.~A.~Cargile\altaffilmark{4,5}, 
A.~Kraus\altaffilmark{6},
N.~M.~Law\altaffilmark{7}, 
J.~J.~Lemonias\altaffilmark{1}, 
H.~G.~Arce\altaffilmark{8}, 
D.~F.~Fierroz\altaffilmark{1}, 
A.~Kundert\altaffilmark{9}
}

\altaffiltext{1}{Columbia University, Department of Astronomy, 550 West 120th Street, New York, NY 10027} 
\altaffiltext{2}{Lowell Observatory, 1400 West Mars Hill Road, Flagstaff, AZ 86001}
\altaffiltext{3}{Haverford College, 370 Lancaster Ave, Haverford, PA 19041}
\altaffiltext{4}{Department of Physics and Astronomy, Vanderbilt University, Nashville, TN 37235}
\altaffiltext{5}{Harvard-Smithsonian Center for Astrophysics, 60 Garden Street, Cambridge, Massachusetts 02138}
\altaffiltext{6}{Department of Astronomy, University of Texas at Austin, 2515 Speedway, Stop C1400, Austin, TX 78712}
\altaffiltext{7}{Department of Physics and Astronomy, University of North Carolina, Chapel Hill, NC 27599}
\altaffiltext{8}{Department of Astronomy, Yale University, New Haven, CT 06520, USA}
\altaffiltext{9}{Department of Astronomy, University of Wisconsin-Madison, 2535 Sterling Hall, 475 North Charter Street, Madison, WI 53706}

\begin{abstract}  
Open clusters are collections of stars with a single, well-determined age, and can be used to investigate the connections between angular-momentum evolution and magnetic activity over a star's lifetime. We present the results of a comparative study of the relationship between stellar rotation and activity in two benchmark open clusters: Praesepe and the Hyades. As they have the same age and roughly solar metallicity, these clusters serve as an ideal laboratory for testing the agreement between theoretical and empirical rotation-activity relations at $\approx$600 Myr. We have compiled a sample of 720 spectra --- more than half of which are new observations --- for 516 high-confidence members of Praesepe; we have also obtained 139 new spectra for 130 high-confidence Hyads. We have also collected rotation periods ($P_{rot}$) for 135 Praesepe members and 87 Hyads. To compare \halpha emission, an indicator of chromospheric activity, as a function of color, mass, and Rossby number $R_o$, we first calculate an expanded set of $\chi$ values, with which we can obtain the \halpha to bolometric luminosity ratio, $L_{H\alpha}/L_{bol}$, even when spectra are not flux-calibrated and/or stars lack reliable distances. Our $\chi$ values cover a broader range of stellar masses and colors (roughly equivalent to spectral types from K0 to M9), and exhibit better agreement between independent calculations, than existing values. Unlike previous authors, we find no difference between the two clusters in their \halpha equivalent width or $L_{H\alpha}/L_{bol}$ distributions, and therefore take the merged \halpha and $P_{rot}$ data to be representative of 600-Myr-old stars. Our analysis shows that \halpha activity in these stars is saturated for $\Ro\leq0.11^{+0.02}_{-0.03}$. Above that value activity declines as a power-law with slope $\beta=-0.73^{+0.16}_{-0.12}$, before dropping off rapidly at $\Ro\approx0.4$. These data provide a useful anchor for calibrating the age-activity-rotation relation beyond 600 Myr.
\end{abstract}

\keywords{stars: rotation, stars: activity, stars: chromospheres, stars: coronae, stars: evolution, stars: late-type}

\section{Introduction}
In \citet[][hereafter Paper~I]{agueros11} we reported stellar rotation periods ($P_{rot}$) for 40 late-K/early-M members of the open cluster Praesepe ($\alpha$ 08 40 24 $\delta$ $+$19 41), also known as the Beehive Cluster, derived from our Palomar Transient Factory \citep[PTF;][]{nick2009, rau2009} observations. By combining these $P_{rot}$ with those obtained by \citet{scholz2007}, \citet{delorme2011}, and \citet{scholz2011}, we determined that Praesepe's mass-period relation transitions from a well-defined singular relation to a more scattered distribution of $P_{rot}$ at $\approx$0.6 \Msun, or a spectral type (SpT) $\approx$M0. We found that the location of this transition is consistent with expectations based on observations of younger clusters and the assumption that stellar spin-down is the dominant mechanism influencing angular momentum evolution at $\approx$600 Myr, the age of Praesepe. 

This mass-period relation is one projection of the relationship between stellar age, rotation, and magnetic activity. Numerous studies of open clusters have derived relationships between a star's age and chromospheric or coronal emission, which are manifestations of magnetic activity \citep[e.g.,][]{skumanich72, radick1987, hawley1999, soderblom2001}. Other studies have used e.g., kinematic information to infer the activity lifetimes of low-mass field stars \citep[e.g.,][]{hawley1999, andy08}. \citet{andy08} model the dynamical heating of stars in the Galactic disk and use the results to calibrate the age-dependence of the vertical gradient in $H\alpha$ emission strengths, finding that the activity lifetimes of stars with SpTs of M2 or later appear to be $>$1 Gyr. Because few active early M stars are observed in the field, the activity lifetimes of M0-M1 stars are less well known, but they are likely $\lapprox$600 Myr \citep[][]{andy08}. Thus, we expect that the boundary between \halpha active and inactive Praesepe members will occur in the M0/M1 spectral range. That this transition occurs at roughly the same mass as that between the singular mass-period relation and a more scattered distribution of $P_{rot}$ strengthens the case for a rotation-activity relation in Praesepe.

\begin{figure*}[!th]
\centerline{\includegraphics[angle=270,width=2\columnwidth]{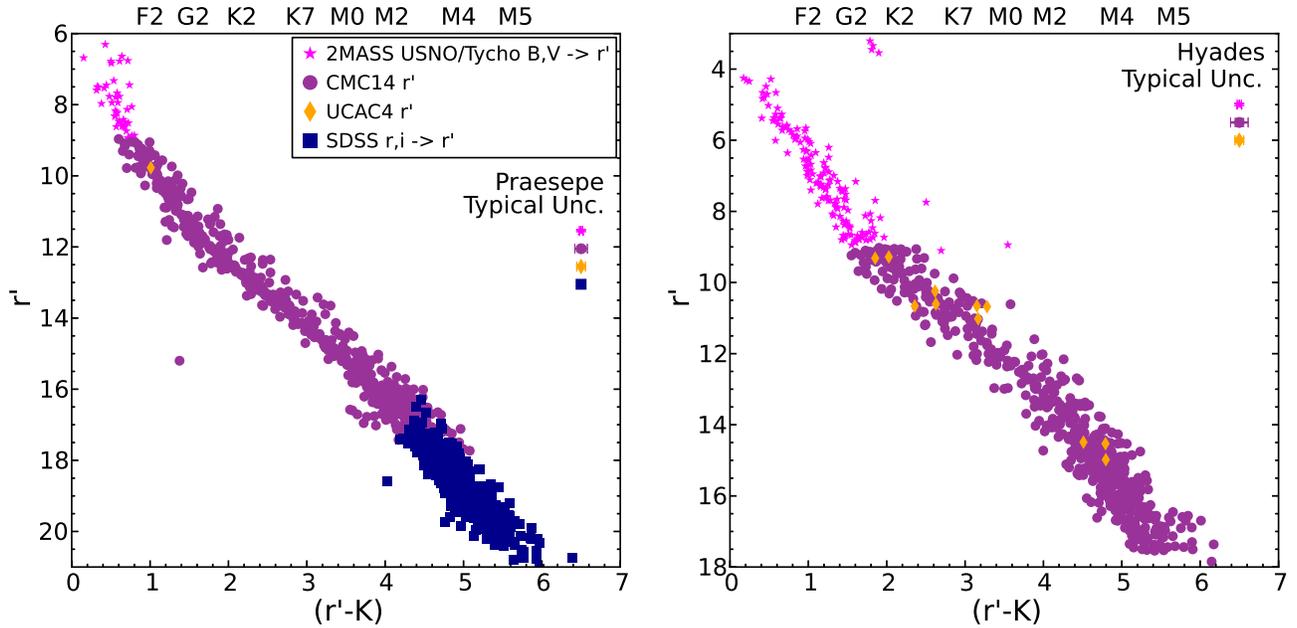}}
\caption{CMD for Praesepe (left) and the Hyades (right) indicating the sources of $r'$ photometry. Typical photometric uncertainties for the surveys used in assembling these CMDs are shown for reference. The y-axes are such that stars of similar masses will appear at roughly the same position, despite the larger distance to Praesepe. SpTs are indicated along the top axes. All the stars from the catalogs discussed in Section~\ref{cats} are shown. While $r'$ is drawn or transformed from multiple surveys, the cluster sequences are clean and well-defined. }
\label{fig:psource} 
\end{figure*} 

In Paper~I, we also compared the mass-period relation for Praesepe to that derived from the rotation data published by \citet{delorme2011} for the Hyades, which is generally assumed to be coeval with Praesepe. This indicated that the transition to a single-valued mass-period relation occurs at a lower characteristic mass in the Hyades, implying that this cluster is older than Praesepe. Intriguingly, activity studies do not necessarily agree with this conclusion: in the most recent large-scale spectroscopic survey of the two clusters, \citet{kafka2006} found that, as measured by \halpha emission strength, more massive stars are active in the Hyades than in Praesepe, implying that the Hyades is the younger cluster. (The coronal activity picture is not much clearer: \citet{franc2003} found that the two clusters have similar X-ray luminosity functions, contradicting the earlier findings of \citet{randich1995}.) 

In this paper we examine activity and rotation in Praesepe and the Hyades to probe the activity-rotation relation at 600 Myr. Our spectroscopic sample includes new spectra obtained with the 2.4-m Hiltner telescope at MDM Observatory, the WIYN 3.5-m telescope at NOAO, both on Kitt Peak, AZ,\footnote{The MDM Observatory is operated by Dartmouth College, Columbia University, Ohio State University, Ohio University, and the University of Michigan. The WIYN Observatory is a joint facility of the University of Wisconsin-Madison, Indiana University, Yale University, and the National Optical Astronomy Observatory.} and the Magellan Echellette (MagE) Spectrograph\footnote{Support for the design and construction of the Magellan Echellette Spectrograph was received from the Observatories of the Carnegie Institution of Washington, the School of Science of the Massachusetts Institute of Technology, and the National Science Foundation in the form of a collaborative Major Research Instrument grant to Carnegie and MIT (AST-0215989).} on the 6.5-m Clay Telescope, Las Campanas, Chile. To these we add spectra from the literature; in total, we have 720 spectra of 516 high-confidence members of Praesepe, and 139 spectra of 130 high-confidence Hyads. We also make use of the Praesepe $P_{rot}$ reported in Paper~I, as well as those measured by \citet{scholz2007}, \citet{delorme2011}, and \citet{scholz2011}. And we supplement the Hyades $P_{rot}$ of \citet{delorme2011} with $P_{rot}$ derived from All Sky Automated Survey \citep[ASAS;][]{ASAS} data by Kundert et al.\ (in prep.). 

We begin in Section~\ref{data} by describing our membership catalogs for both clusters, the sources of our photometric data, our spectroscopic sample, and our collection of $P_{rot}$ data. In Section~\ref{meas}, we describe our method for measuring \halpha equivalent widths (EqWs) and for deriving the ratio of the \halpha line luminosity over the stellar bolometric luminosity ($L_{H\alpha}/L_{bol}$). We also discuss our procedure for calculating masses, identifying binaries, and determining Rossby numbers ($R_o$). In Section~\ref{res}, we use  our \halpha data to compare chromospheric activity in the two clusters and present an updated 600-Myr mass-period relation that includes data for both clusters. We then examine the relation between \halpha emission and rotation, and between X-ray emission and rotation, for stars in our sample. We conclude in Section~\ref{concl}. 

Our $\chi$ values were calculated as a function of color using medium-resolution synthetic spectra, and as a function of SpT using field M dwarfs.
 As they differ from those of \citet{walkowicz2004} and \citet{west2008}, in Appendix~\ref{appen} we discuss in greater detail our calculations and provide tables of our $\chi$ values. 

\section{Data}\label{data}
\subsection{Membership Catalogs}\label{cats}
\citet{adam2007} calculated proper motions and photometry for several million objects within 7\degree\ of the center of Praesepe. The resulting catalog includes 1128 candidate cluster members with membership probabilities $P_{mem} > 50$\%.  As in Paper I, we supplement this catalog with 41 known members that are too bright to be identified as members by \citet{adam2007}. 

For the Hyades, we adopt the \citet{roser2011} membership catalog. These authors identified candidate Hyades members via the convergent point method and confirmed membership using photometry. The \citet{roser2011} catalog does not include $P_{mem}$ calculations, but the authors list contamination percentages based on distance from the cluster center ($d_c$): the contamination is $1\%$ for stars with $d_c\le9$~pc, $7.5\%$ for $9 <d_c\le18$~pc, and $30\%$ for $18 <d_c\le30$~pc. We converted these to $P_{mem}$ by subtracting the contamination percentage from $100\%$. By our calculations, the catalog includes 724 stars with $M \ge 0.12~\Msun$ and $P_{mem} \ge 70\%$ up to 30 pc from the cluster center. Based on photometric limits, \citet{roser2011} state that their catalog is complete down to $\approx$0.25~\Msun.

We supplement the \citet{roser2011} catalog with new Hyades members found by Kundert et al.\ (in prep.), who identify 170 cluster members based on reduced proper motions ($\mu$) and distances obtained by {\it Hipparcos} \citep{hip}. Kundert et al.\ consider stars within 26\degree\ and 20 pc of the cluster center and with $-170<\mu_{\parallel}<-60$ and $-20<\mu_{\perp}<20$ mas~yr$^{-1}$. All but 13 of the Hyades members identified in this manner were also identified by \citet{roser2011}. We add these 13 additional members to our catalog, bringing the total number of Hyads to 737.

\subsection{Photometry}\label{photometry}
We use $(r'-K)$ as our primary proxy for stellar temperature. By selecting an optical-NIR color, we obtain a broader dynamic range than is possible with a narrower color index. For example, in $(J-K)$, M-dwarf colors range from roughly $0.9~\textrm{to}~1.2$ mag, while this same mass range is spread out from $3.3~\textrm{to}~8.0$ mag in $(r'-K)$. While nearly all the stars in our sample have Two Micron All Sky Survey \citep[2MASS;][]{2mass} $K$-band magnitudes, the large range in $r'$ magnitudes ($\approx$15 mag) for both clusters meant that we had to obtain this photometry from multiple sources.

The Carlsberg Meridian Catalogue 14 \citep[CMC14;][]{cmc14} provides photometry for approximately $10^8$ stars with declinations between $-30$\degree\ and $50$\degree\ and $9\ < r'\ \lapprox\ 17$ mag. We use CMC14 photometry for stars falling within this magnitude range. The 4$^{\rm th}$ U.S.~Naval Observatory CCD Astrograph Catalog \citep[UCAC4;][]{zacharias2012} includes $g'r'i'$ magnitudes from APASS \citep{apass}.  In the Hyades, we use the CMC14 magnitudes and errors listed in \cite{roser2011}.  CMC14 does not list $r'$ errors for all stars in Praesepe; in these cases we use the typical errors for the catalog.\footnote{\url{http://www.ast.cam.ac.uk/ioa/research/cmt/cmc14.html}} For a handful of stars with $10\ \lapprox\ r'\ \lapprox\ 14$ mag that do not appear in CMC14, we use $r'$ magnitudes from UCAC4.

For stars lacking $r'$ magnitudes, we use the \citet{jester2005} and \citet{bilir2008} transformations to convert available $r$ magnitudes to $r'$.\footnote{We convert $r$ to $r'$ rather than the inverse because CMC14 lacks the $i'$ photometry that would allow us to transform $r'$ into $r$. Furthermore, the \citet{bilir2008} relation for $(r-i)$ as a
function of 2MASS colors is valid for $(r-i)\ \lapprox\ 0.5$, and we could only apply it to the highest mass dwarfs in these clusters. The difference between $r$ and $r'$ is small
but not negligible for our purposes.}

For stars in Praesepe, we use SDSS photometry to obtain $r'$ for stars with $r > 16$ mag. Our $(r-K)$ color-magnitude diagram (CMD) in Paper~I indicated that the SDSS $r$ magnitudes could not always be trusted for stars brighter than $r \approx 16$, even in cases where the SDSS flags did not indicate that the star was saturated (see figure 4 in Paper~I). We use SDSS $ri$ photometry to obtain $r'$ by applying the  \citet{jester2005} equations given on the
SDSS website.\footnote{\url{http://www.sdss.org/dr7/algorithms/jeg\_photometric\_eq\_dr1.html}} Few Hyads are in the SDSS footprint, and many of those in the footprint are saturated; as a result, we do not use any SDSS magnitudes for Hyads.

For stars with $r<9$ mag in both clusters, we use the \citet[][]{jester2005} relations to convert the USNO-A2.0 and Tycho 2 Johnson $B$ and $V$ magnitudes included in the 2MASS catalog to SDSS $r$ magnitudes. Since these stars fall into the appropriate color range, we then apply the \citet{bilir2008} transformation from 2MASS colors to obtain $(r-i)$ for these stars. Finally, we use these $r$ and $(r-i)$ values to obtain $r'$ applying the \citet{jester2005} relation, as above. Figure~\ref{fig:psource} shows the $r'$ versus~$(r'-K)$ CMDs for both clusters. The typical photometric uncertainty for these $r'$ magnitudes depends on the source catalog; after applying the conversions discussed above to 2MASS or SDSS data, the uncertainty is generally $\lesssim$0.1 mag. For CMC14 data, the uncertainty is $\approx$0.1 mag for Hyads and slightly smaller for stars in Praesepe; for UCAC4 data, it is $\approx$0.05 mag.

\setlength{\tabcolsep}{2.5pt}
\begin{deluxetable}{lcc}[!h]
\tablewidth{0pt}
\tabletypesize{\scriptsize}
\tablecaption{ModSpec Observations of Praesepe and Hyades Stars \label{mdmstats}}
\tablehead{
\colhead{}     & \multicolumn{2}{c}{\# of Spectra}    \\ 
\colhead{Dates} & \colhead{Praesepe} & \colhead{Hyades}  
}
\startdata
2010 Dec 02-Dec 06 & 124 & \nodata  \\ 
2011 Feb 08-Feb 11 & 82  & \nodata \\ 
2011 Nov 30-Dec 05 & \nodata   & 66 \\ 
2012 Feb 17-Feb 21 & 44  & 13 \\ 
2012 Nov 11-Nov 14 & 8 & 65 \\
\hline
Total	& 258	& 144  
\enddata
\tablecomments{All dates in Tables~\ref{mdmstats}-\ref{magestats} are UT.}
\end{deluxetable}

\begin{figure*}[!th]
\centerline{\includegraphics[angle=270,width=2\columnwidth]{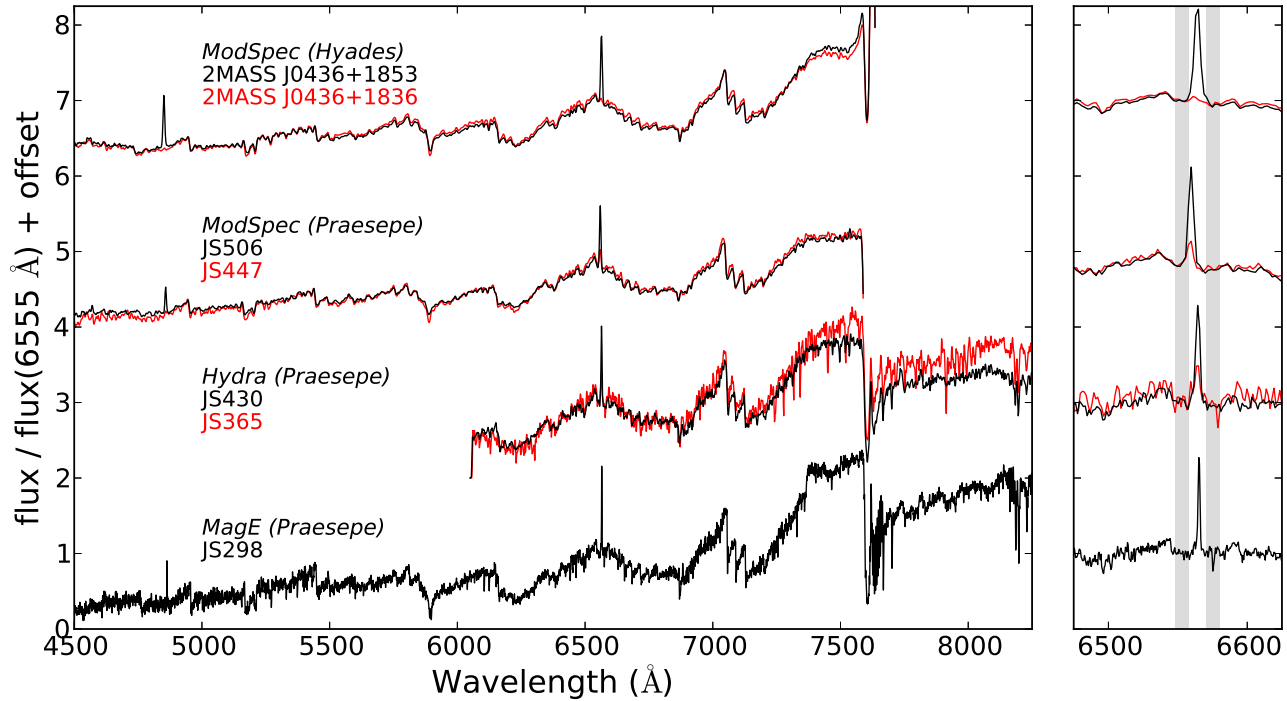}}
\caption{Example spectra from our new observations of Praesepe and Hyades stars with ModSpec on MDM, Hydra on WIYN, and MagE on Magellan. The Hyads are both M2 stars, while all the Praesepe stars are M3s. For each cluster/instrument combination, the most active star of that SpT is shown in black, and the least active in red. The panel to the right shows a close-up of the region around $H\alpha$, with gray shading marking the typical continuum regions for the EqW measurements (see Section~\ref{activity}).} 
\label{fig:specs}
\end{figure*} 

\subsection{Spectroscopy}\label{spectroscopy}
\subsubsection{New Observations}
We used the MDM Observatory Modular Spectrograph (ModSpec) on the Hiltner 2.4-m telescope to obtain spectra of stars in Praesepe and the Hyades over the course of five multi-night runs between 2010 Dec 2 and 2012 Nov 14 (see Table~\ref{mdmstats}). ModSpec was configured to provide coverage from 4500 to 7500~\AA~with $\approx$1.8~\AA~sampling and a spectral resolution of $\approx$3300. These spectra were reduced with a script written in PyRAF, the Python-based command language for the Image Reduction and Analysis Facility (IRAF).\footnote{PyRAF is a product of the Space Telescope Science Institute, which is operated by AURA for NASA. IRAF is distributed by the National Optical Astronomy Observatories, which are operated by the Association of Universities for Research in Astronomy, Inc., under cooperative agreement with the National Science Foundation.} All the spectra were trimmed, overscan- and bias-corrected, cleaned of cosmic rays, flat-fielded, extracted, dispersion-corrected, and flux-calibrated using standard IRAF tasks. After accounting for the quality of the spectra and for those stars that we observed more than once, our sample included 253 spectra for 209 Praesepe stars, of which 226 spectra were for 187 stars with $P_{mem} > 70\%$, as calculated by \citet{adam2007}. Our Hyades sample included 139 spectra for 130 stars with $P_{mem} > 70\%$ (see Section~\ref{cats}) once the same quality cuts were made.

\setlength{\tabcolsep}{2.5pt}
\begin{deluxetable}{lccc}[!h]
\tablewidth{0pt}
\tabletypesize{\scriptsize}
\tablecaption{Hydra Observations of Praesepe Fields \label{hydrastats}}
\tablehead{
\colhead{}     & \colhead{} & \colhead{Exposure} & \colhead{\# of} \\
\colhead{Date} & \colhead{Field Center} & \colhead{Time (s)}  & \colhead{Spectra}
}
\startdata
2011 Feb 7     & 08 39 22.3 $+$20 02 00.0 & $1380$ & 57 \\ 
    & 08 40 24.0 $+$19 36 00.0 & $6000$ & 41 \\ 
    & 08 39 07.5 $+$20 44 00.0 & $6000$ & 24 \\ 
    & 08 45 19.0 $+$19 18 00.0 & $4200$ & 26 \\ 
    & 08 41 51.5 $+$19 30 00.0 & $1500$ & 43 \\ 
2011 Feb 8     & 08 39 07.5 $+$20 44 00.0 & $4200$ & 23 \\ 
    & 08 44 35.5 $+$20 12 00.0 & $3600$ & 17 \\
\hline
Total		&	&	& 231
\enddata
\end{deluxetable}

\setlength{\tabcolsep}{2.5pt}
\begin{deluxetable}{llcc}[!h]
\tablewidth{0pt}
\tabletypesize{\scriptsize}
\tablecaption{MagE Observations of Praesepe Stars \label{magestats}}
\tablehead{
\colhead{}     & \colhead{}     & \colhead{} & \colhead{Exposure}  \\
\colhead{Date} & \colhead{Target} & \colhead{Position} & \colhead{Time (s)} 
}
\startdata
2011 Mar 19 & JS 718\tablenotemark{a} & 08 40 04.2 $+$19 24 50.3 & 1600 \\
	    & HSHJ 428 & 08 42 37.6 $+$19 59 18.9 & 1800 \\
2011 Mar 20 & JS 123\tablenotemark{b} & 08 36 19.2 $+$19 53 54.9 & 900 \\
	    & JS 298 & 08 39 31.8 $+$19 24 17.6 & 1200 \\
	    & JS 729 & 08 41 26.0 $+$19 59 15.1 & 900
\enddata
\tablenotetext{a}{Identified as a candidate binary system in Paper~I.}
\tablenotetext{b}{Identified as a candidate binary system in this paper.}
\end{deluxetable}

\setlength{\tabcolsep}{2.5pt}
\begin{deluxetable}{lccc}
\tablewidth{0pt}
\tabletypesize{\scriptsize}
\tablecaption{Final Spectroscopic Sample \label{specstats}}
\tablehead{
\colhead{Telescope} & \colhead{Hiltner}  & \colhead{WIYN}   
& \colhead{Magellan}\\
\colhead{(Instrument)} &  \colhead{(ModSpec)}  & \colhead{(Hydra)}
& \colhead{(MagE)}
}
\startdata
Praesepe stars        		& 209 & 176 & 5\\
... with $P_{mem} > 70$\%    	& 187 & 174 & 5 \\
... with spectra in literature\tablenotemark{a}	& 42  & 61  & 4 \\
Hyades stars             	& 130 & \nodata & \nodata \\
... with $P_{mem} > 70$\% 	& 130 & \nodata & \nodata 
\enddata
\tablenotetext{a}{These are for the stars with $P_{mem} > 70\%$. See Section~\ref{arch}.}
\end{deluxetable}

\begin{figure*}[!th]
\centerline{\includegraphics[angle=270,width=2.\columnwidth]{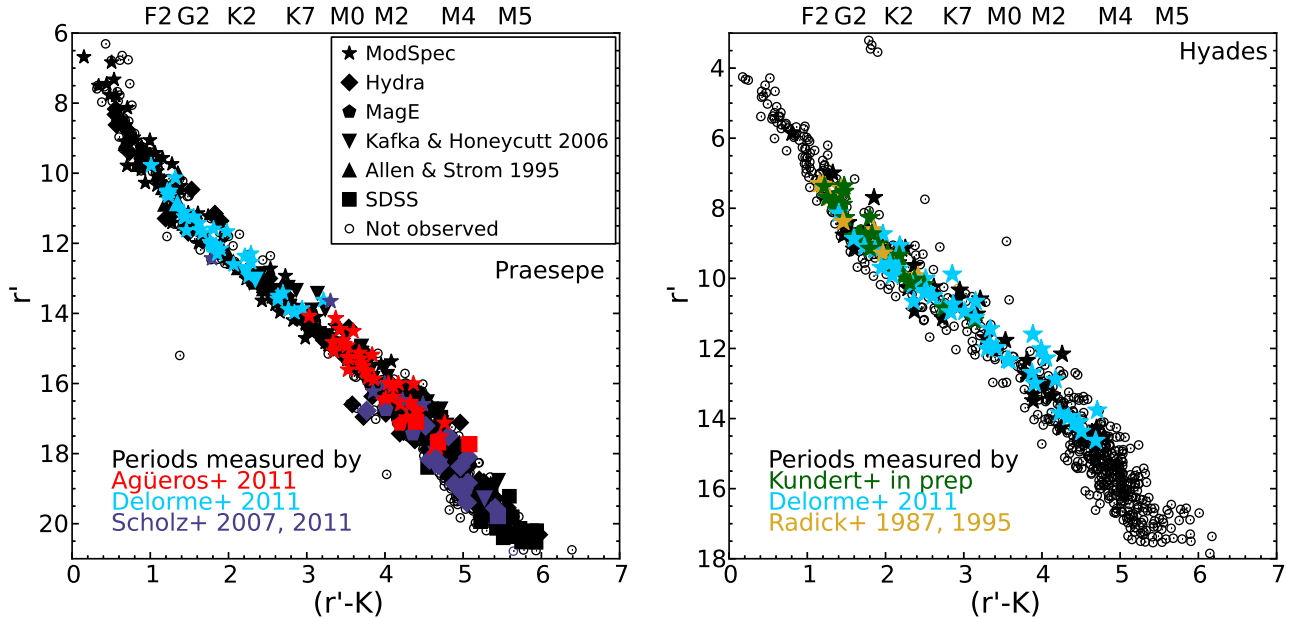}}
\caption{CMD for Praesepe (left) and the Hyades (right) showing the completeness of our spectroscopic samples. We have obtained spectra across the full mass range for which $P_{rot}$ have been measured in each cluster.} 
\label{fig:completeness} 
\end{figure*} 

We also observed Praesepe with the Hydra multi-object spectrograph on the WIYN 3.5-m telescope during the nights of 2011 Feb 7 and~8. We used the bench-mounted spectrograph with the red fiber cable and an \'echelle grating with 600 lines~mm$^{-1}$ set at a blaze angle of 13.9\degree. This resulted in coverage from 6050 to 8950~\AA~with $\approx$1.4~$\AA$ sampling and a spectral resolution of $\approx$4000. We targeted seven fields that required exposure times ranging from $1380$ to $6000$ s and were typically divided into three or four sub-exposures for cosmic-ray removal (see Table~\ref{hydrastats}). We reduced these spectra using standard routines in the IRAF Hydra package.\footnote{\url{http://iraf.noao.edu/tutorials/dohydra/dohydra.html}} Each image was trimmed and instrument biases were removed. The spectra for the individual fibers were extracted, flat-fielded, and dispersion-corrected. Sky spectra from $\approx$30 fibers placed evenly across the field-of-view were combined and subtracted from our target star spectra. We throughput-corrected and flux-calibrated each spectrum using the flux standard G191B2B, which was observed using the same set-up as for our targets. We then combined the sub-exposures for each object to form a high signal-to-noise (S/N) spectrum for each star.

We placed Hydra fibers on 231 targets in Praesepe; 43 of these spectra were too noisy to use for our analysis, so that the total number of usable spectra was 188. Once we accounted for the stars targeted more than once, there were a total of 176 individual Praesepe members with at least one usable Hydra spectrum. 174 of these stars have $P_{mem} > 70\%$ in the \citet{adam2007} catalog, and we have a total of 186 Hydra spectra for these stars. (We observed 23 stars with both ModSpec and Hydra.) 

Finally, five Praesepe rotators were observed with MagE \citep[][]{mage} on the 6.5-m Clay Telescope on the nights of 2011 Mar 19 and Mar 20 (see Table~\ref{magestats}). MagE is a cross-dispersed spectrograph that covers 3000$-$10500~$\AA$ in a single exposure. These spectra were reduced with the MASE pipeline \citep{mage2}. All five stars have $P_{mem} > 70\%$. 

Example spectra from each observatory are shown in Figure \ref{fig:specs}; Table~\ref{specstats} provides the overall statistics for our spectroscopic campaign, and reflects the application of the quality cuts discussed above to the data. In Praesepe, our goal was to obtain spectra for at least twice as many stars of a given SpT without measured periods as for stars with known periods, and we achieved this for stars later than K4. In the Hyades, by contrast, we mostly observed stars with known periods.

\subsubsection{Archival Spectroscopy}\label{arch}
To increase our spectroscopic coverage of Praesepe, we collected spectra from the literature. \citet{allen1995} compiled a grid of stellar classification spectra using Hydra on the Mayall 4-m telescope at NOAO, Kitt Peak, and observed 98 stars classified by \citet{adam2007} as Praesepe members. (They also observed four non-members.) These spectra were flat-fielded and wavelength-calibrated, but were not flux-calibrated and have no associated noise spectrum. We removed two spectra from this sample because they were too noisy for our purposes. Of the remaining spectra, 93 are for stars with $P_{mem} > 70\%$.

\citet{kafka2004,kafka2006} observed 224 K and M dwarfs in Praesepe using Hydra on the WIYN 3.5-m telescope. S.~Kafka (pers.~comm.) kindly provided us with 185 of these spectra, which are not flux-calibrated or corrected for telluric absorption. After visual inspection, we removed 24 spectra due to incomplete cosmic-ray subtraction and/or strong sky lines near $H\alpha$. Of the remaining spectra, 154 are for stars with $P_{mem} > 70\%$. 

As of 2013 Feb 14, SDSS had obtained spectra for 66 Praesepe stars. These spectra have been sky-subtracted, corrected for telluric absorption, and spectrophotometrically calibrated, as well as calibrated to heliocentric vacuum wavelengths.\footnote{\url{http://www.sdss.org/dr3/products/spectra/}} We removed two spectra from this sample because they were too noisy for our purposes; 56 of the remaining spectra are for stars with $P_{mem} > 70\%$.

For the Hyades, J.~Stauffer (pers.~comm.) kindly shared with us 12 spectra obtained as part of the \citet{stauffer1997} survey of the cluster. These spectra, along with 161 of the Praesepe spectra shared with us by S.~Kafka, were used to test our EqW measurements against those in the literature (see Section~\ref{activity}).

Once the quality cuts described above and the $P_{mem} = 70\%$ threshold was set, and we accounted for stars with multiple spectra, we were left with 720 spectra of 516 Praesepe members and 139 spectra of 130 Hyads. Figure~\ref{fig:completeness} gives an overview of our spectral coverage in each cluster, along with the distribution of stars with $P_{rot}$ measured by the surveys discussed below. 

\subsection{Rotation Periods}\label{prot}
The Palomar Transient Factory is described in detail in \citet{nick2009} and \citet{rau2009}; our first season of PTF observations of Praesepe and subsequent light-curve analysis is described in Paper~I. This analysis produced high-confidence measurements of $P_{rot}$ ranging from 0.52 to 35.85 d for 40 stars. Thirty-seven of these stars have $P_{mem} > 95\%$, as calculated by \citet{adam2007}, with two of the other stars having $P_{mem} > 94\%$. 

In Paper~I, we also compiled $P_{rot}$ measurements from the literature, including 52 bright stars (of which 46 have $P_{mem} > 95\%$), whose periods were measured by \citet{delorme2011}, and 54 low-mass Praesepe members with periods reported by \citet{scholz2007} and \citet{scholz2011}. As nine of these stars with $P_{rot}$ from the literature also have PTF periods, the total sample of Praesepe rotators is 135 stars. Our spectroscopic sample includes observations of 113 of these stars, of which 111 have $P_{mem} > 70\%$. 

\citet{radick1987, radick1995} searched for variability in Hyades stars using differential photometry obtained over several seasons, at least one of which had a five-month baseline.  These authors measured $P_{rot}$ for 18 cluster members, all with SpT K8 or earlier.  

In addition to their results for Praesepe, \citet{delorme2011} published 60 $P_{rot}$ for Hyades stars that were also derived from data collected by the SuperWASP search for transiting exoplanets. \citet{delorme2011} analyzed light curves spanning $\gtrsim$100 d for stars within $\approx$15\degree\ of the Hyades's center. Fifty-nine of their rotators have $P_{mem} > 95\%$ according to their analysis.  

Kundert et al.~(in prep) used the publicly available light curves from ASAS \citep{ASAS} to measure $P_{rot}$ for Hyades stars. On average, the ASAS data provide 240 observations over a seven-year baseline for $V$ = 7$-$13 mag stars. Kundert et al.~measure $P_{rot}$ for 40 Hyads; 18 are new measurements. For the other 22, the agreement with the $P_{rot}$ measured by \cite{radick1987,radick1995} and \cite{delorme2011} is excellent, with the exception of ASAS 040526$+$1926.5. For this star, Kundert et~al.~find a $P_{rot}$ half that published by \cite{delorme2011}; we use this more recent period for our analysis.

Nine Hyades rotators are known binaries, and we remove these stars from the list of rotators for our analysis. There are no known binaries among the Praesepe rotators (see Section~\ref{binaries}). This leaves 87 known rotators in the Hyades, and we have spectra for 83 of those stars. 

\section{Measurements and Derived Quantities}\label{meas}
\subsection{\halpha Measurements and $L_{H\alpha}/L_{bol}$}\label{activity}
We measured the equivalent width (EqW) of the \halpha line for each spectrum in our sample.  We did not correct these measurements for photospheric absorption. Where possible, the continuum flux was taken to be the average flux between 6550$-$6560~\AA\ and 6570$-$6580~\AA\ (as shown in the right panel of Figure~\ref{fig:specs}). In cases where the line was broad or shifted away from 6563~\AA, the continuum flux was measured from 10~\AA\ windows on each side of the line. The window used to measure the line flux varies from spectrum to spectrum, and was adjusted interactively. 

In cases where we had multiple spectra for a star, the EqWs were generally consistent at the $1\sigma$ level. A few stars appeared to show strongly varying $H\alpha$ emission. We have spectral coverage blueward of $H\alpha$ for a small number of these stars, and these do not appear to be flaring. We therefore simply use the average EqW in all these cases for our analysis.

To estimate the EqW uncertainties, the same person first measured each EqW twice, and we took the difference between the two measurements to be the human error in the interactive measurement.  The median difference between the two measurements was $0.22~\AA$ in Praesepe and $0.15~\AA$ in the Hyades.  

\begin{figure}
\centerline{\includegraphics[width=1.0\columnwidth]{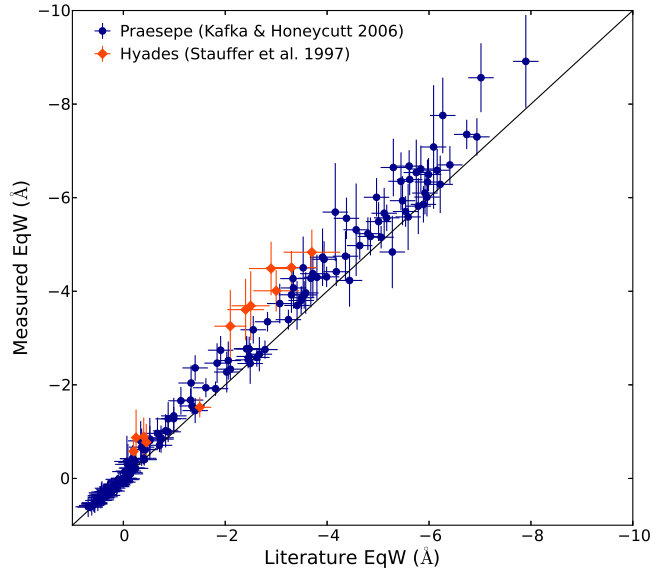}}
\caption{Comparison of EqW measurements for the 161 \citet{kafka2006} spectra of Praesepe stars and for 12 \citet{stauffer1997} spectra of Hyades stars. We follow the convention that an EqW $<$ 0 corresponds to emission. While our EqW measurements are consistent with those in the literature for these spectra, our EqWs are systematically larger, and the difference grows as the EqWs become larger.} 
\label{fig:compprae} 
\end{figure}

We then used a Monte Carlo technique to add Gaussian noise to each point in the spectrum and remeasured the EqWs 2500 times in an automated fashion. The continuum and line regions from the initial interactive measurements were re-used. For spectra with an associated uncertainty spectrum, we drew the noise at each point from a Gaussian with width equal to the uncertainty at that point. For stars without an uncertainty spectrum, we drew the noise from a Gaussian distribution with a width equal to the $\sigma$ of the flux in the continuum region.  We took the standard deviation of the EqWs from the Monte Carlo simulation as the error from noise in the spectrum.  The two error measurements were added in quadrature to produce the uncertainty in each EqW.  

{In Figure~\ref{fig:compprae}, we compare our EqW measurements to those of \citet{kafka2006} for all 161 of the usable Praesepe spectra collected by these authors, and to those of \citet{stauffer1997} for 12 spectra of Hyads collected by those authors. \citet{kafka2006} state that their typical EqW uncertainties are $\approx$0.2~$\AA$. \citet{stauffer1997} do not quote a typical EqW uncertainty, but we assume a $15\%$ measurement uncertainty, as quoted by \citet{stauffer1994} in earlier work on the Hyades. While the measurements are broadly consistent with each other, our EqW measurements tend to return values 10$-$20\% larger than those of \citet{kafka2006} and 25$-$35\% larger than those of \citet{stauffer1997}.

The Hyades EqW measurements used by \citet{kafka2006} come from a variety of sources and therefore potentially very different telescope/spectrograph combinations and EqW-measurement techniques. As shown in Figure~\ref{fig:compprae}, the difference between our measurements and those of \citet{stauffer1997} is larger than the difference between our measurements and those of \citet{kafka2006}. These discrepancies argue for a single, uniform approach to measuring EqWs, as is possible for our spectroscopic sample, to maximize the internal consistency of the results.

Figure~\ref{fig:eqw} shows the average EqW for all stars in our sample with $P_{mem}\ge70\%$;  these values can be found in Tables~\ref{prae} (for Praesepe) and \ref{hyad} (for the Hyades). 

The ratio of the \halpha luminosity to the bolometric luminosity of the star, $L_{H\alpha}/L_{bol}$, enables a better comparison of activity between stars of different (low) masses than EqW alone. It reflects the importance of the \halpha flux relative to the star's entire energy output, and not just relative to the continuum flux in a single band, which changes rapidly across the K and M SpTs \citep[][]{nlds}. Ideally, $L_{H\alpha}/L_{bol}$ would be calculated as
\begin{eqnarray*}
\frac{L_{H\alpha}}{L_{bol}} = -W_{H\alpha} \frac{f_0}{f_{bol}}
\end{eqnarray*}
where $W_{H\alpha}$ is the EqW of the \halpha line, $f_0$ is the continuum flux for the line, and $f_{bol}$ is the apparent bolometric flux of the star. However, because some of our spectra are not flux-calibrated, we cannot always measure $f_0$ directly. We therefore followed \citet{walkowicz2004} and \citet{west2008} in calculating $\chi = {f_0}/{f_{bol}}$ as a function of color. 

As our sample of active stars includes K and M dwarfs, we could not simply use the $\chi$ values of \citet{walkowicz2004}, which were calculated for M0.5-L0 stars. We therefore calculated $\chi$ as a function of color and magnitude using medium-resolution model spectra from PHOENIX ACES atmospheres \citep{husser2013}; we obtained synthetic photometry by convolving these spectra with the SDSS and 2MASS filter curves.\footnote{We calibrated $\chi$ as a function of color rather than absolute magnitude because the distances to many of the low-mass Hyads have not been directly determined, and the cluster's extent along the light-of-sight is large enough to introduce significant uncertainties in the luminosities.} Interestingly, our $\chi$ values do not match those given in \citet{walkowicz2004} and \citet{west2008}; see Appendix~\ref{appen} for full discussion. Our $\chi$ values are listed as a function of temperature and color in Table~\ref{phoenix}.

\begin{deluxetable*}{lccr@{ $\pm$ }lccccr@{ $\pm$ }lr@{ $\pm$ }l}[!h]
\tablewidth{0pt}
\tabletypesize{\scriptsize}
\tablecaption{Praesepe Stars \label{prae}}
\tablehead{
\colhead{Name}  &  \colhead{RA}  &  \colhead{DEC}   &  \multicolumn{2}{c}{$r'$}  & \colhead{$r'$ src\tablenotemark{$\dagger$}} & \multicolumn{1}{c}{M}  & \colhead{\Ro}  &  \colhead{Binary?} & \multicolumn{2}{c}{$H\alpha$ EqW}  &  \multicolumn{2}{c}{$L_{H\alpha}/L_{bol}$\tablenotemark{$\ddagger$}}\\ 
\colhead{}  &  \colhead{}  &  \colhead{}  &  \multicolumn{2}{c}{(mag)}  &  \colhead{}  &  \colhead{(\Msun)}  &  \colhead{} & \colhead{}  & \multicolumn{2}{c}{($\AA$)}  & \multicolumn{2}{c}{($\times10^{-5}$)}
}
\startdata
JS 466   & 08:41:58.84 & 20:06:27.1 & $12.818$ & $0.002$ & C & 0.80 & 0.5525 & N  & $0.560$ & $0.223$ & \multicolumn{2}{c}{\nodata} \\ 
JS 468   & 08:41:59.35 & 19:44:45.1 & $15.631$ & $0.070$ & C & 0.50 & 0.4641 & N  & $-0.300$ & $0.435$ & $1.363$ & $1.984$ \\ 
HSH J404 & 08:42:01.59 & 19:26:46.0 & $17.878$ & $0.005$ & S & 0.22 & \nodata & N & $-2.098$ & $0.498$ & $6.101$ & $1.734$ \\ 
AD 3050 & 08:42:04.48 & 19:32:42.7 & $18.834$ & $0.005$ & S  & 0.19 & 0.0086 & N  & $-8.071$ & $0.619$ & $17.136$ & $3.235$ \\ 
AD 3051 & 08:42:04.69 & 19:38:00.8 & $20.784$ & $0.008$ & S  & 0.13 & 0.0060 & N  & \multicolumn{2}{c}{\nodata} & \multicolumn{2}{c}{\nodata} \\ 
JS 470   & 08:42:05.17 & 20:57:56.5 & $15.466$ & $0.070$ & C & 0.58 & \nodata & Y & $-0.456$ & $0.207$ & $1.853$ & $0.876$ \\ 
KW 445   & 08:42:06.49 & 19:24:40.4 & $7.944$ & $0.001$ & T  & 2.38 & \nodata & Y & \multicolumn{2}{c}{\nodata} & \multicolumn{2}{c}{\nodata} \\ 
HSH J421 & 08:42:23.82 & 19:23:12.5 & $18.370$ & $0.006$ & S & 0.21 & 0.0034 & N  & $-4.488$ & $0.507$ & $10.121$ & $2.076$ \\ 
HSH J424 & 08:42:30.77 & 19:29:31.0 & $17.610$ & $0.005$ & S & 0.23 & \nodata & N & $-1.036$ & $0.503$ & $3.326$ & $1.687$ \\ 
JS 513   & 08:43:05.28 & 19:27:54.6 & $14.765$ & $0.035$ & C & 0.58 & \nodata & N & $0.139$ & $0.108$ & \multicolumn{2}{c}{\nodata}
\enddata
\tablenotetext{$\dagger$}{Source of $r'$ magnitude: T is 2MASS/TYCHO2, U is UCAC4, S is SDSS, and C is CMC14.}
\tablenotetext{$\ddagger$}{Only for stars with \halpha in emission.}
\tablecomments{Only selected lines are shown.  The full table, with additional data columns, may be found in the on-line edition of the journal.}
\end{deluxetable*}

\begin{deluxetable*}{lccr@{ $\pm$ }lcccccr@{ $\pm$ }lr@{ $\pm$ }l}[!t]
\tablewidth{0pt}
\tabletypesize{\scriptsize}
\tablecaption{Hyades Stars \label{hyad}}
\tablehead{
\colhead{2MASS J}  &  \colhead{RA}  &  \colhead{DEC}   &  \multicolumn{2}{c}{$r'$}  & \colhead{$r'$ src\tablenotemark{$\dagger$}} & \multicolumn{1}{c}{M}  & \colhead{\Ro}  &  \colhead{Binary?} & \colhead{SpT\tablenotemark{$\star$}} & \multicolumn{2}{c}{$H\alpha$ EqW}  &  \multicolumn{2}{c}{$L_{H\alpha}/L_{bol}$\tablenotemark{$\ddagger$}}\\ 
\colhead{}  &  \colhead{}  &  \colhead{}  &  \multicolumn{2}{c}{(mag)}  &  \colhead{}  &  \colhead{(\Msun)}  &  \colhead{} & \colhead{} & \colhead{} & \multicolumn{2}{c}{($\AA$)}  & \multicolumn{2}{c}{($\times10^{-5}$)}
}
\startdata
03014830$+$3733202 & 03:01:48.32 &37:33:20.3 & $15.12$ & $ 0.12$ & C  & 0.20 & \nodata & N & M4 & $-4.533$ & $0.897$ & $12.070$ & $3.058$ \\ 
03550142$+$1229081 & 03:55:01.36 &12:29:08.2 & $9.74$ & $ 0.04$ & C  & 0.63 & 0.429 & N & K5 & $0.706$ & $0.030$ & \multicolumn{2}{c}{\nodata} \\ 
03550647$+$1659545 & 03:55:06.41 &16:59:54.7 & $8.74$ & $ 0.00$ & T  & 0.99 & 0.664 & N & K1 & $1.058$ & $0.103$ & \multicolumn{2}{c}{\nodata} \\ 
04070122$+$1520062 & 04:07:01.15 &15:20:06.3 & $10.03$ & $ 0.04$ & C  & 0.69 & 0.581 & N & K7 & $0.586$ & $0.153$ & \multicolumn{2}{c}{\nodata} \\ 
04070323$+$2016510 & 04:07:03.25 &20:16:50.9 & $15.58$ & $ 0.18$ & C  & 0.16 & \nodata & N & \nodata & \multicolumn{2}{c}{\nodata} & \multicolumn{2}{c}{\nodata} \\ 
04084015$+$2333257 & 04:08:40.18 &23:33:25.6 & $12.34$ & $ 0.04$ & C  & 0.59 & \nodata & Y & M2 & $-0.510$ & $0.430$ & $2.235$ & $1.908$ \\ 
04142562$+$1437300 & 04:14:25.59 &14:37:30.3 & $8.27$ & $ 0.00$ & T  & 1.03 & \nodata & N & F9 & $1.692$ & $0.030$ & \multicolumn{2}{c}{\nodata} \\ 
04151038$+$1423544 & 04:15:10.34 &14:23:54.6 & $10.96$ & $ 0.04$ & C  & 0.48 & 0.370 & N & K7 & $0.333$ & $0.167$ & \multicolumn{2}{c}{\nodata} \\ 
04322565$+$1306476 & 04:32:25.59 &13:06:47.8 & $10.58$ & $ 0.04$ & C  & 0.97 & \nodata & N & M0 & $-1.728$ & $0.387$ & $11.650$ & $2.858$ \\ 
04343992$+$1512325 & 04:34:39.94 &15:12:32.6 & $11.77$ & $ 0.04$ & C  & 0.61 & \nodata & Y & M1 & $-1.790$ & $0.463$ & $9.048$ & $2.575$ 
\enddata
\tablenotetext{$\dagger$}{Source of $r'$ magnitude: T is 2MASS/TYCHO2, U is UCAC4, and C is CMC14.}
\tablenotetext{$\star$}{SpTs are from the output of the Hammer.}
\tablenotetext{$\ddagger$}{Only for stars with \halpha in emission.}
\tablecomments{Only selected lines are shown.  The full table, with additional data columns, may be found in the on-line edition of the journal.}
\end{deluxetable*}
We then computed $L_{H\alpha}/L_{bol}$ for stars with \halpha in emission using our EqWs, each star's $(r'-K)$, and the appropriate $\chi$ value from our $\chi$ versus~$(r'-K)$ relation. We also calculated $2\sigma$ EqW upper limits for all stars with EqWs consistent with absorption at the $1\sigma$ level, and converted those upper limits into  $L_{H\alpha}/L_{bol}$ upper limits. (The $L_{H\alpha}/L_{bol}$ values can also be found in Tables~\ref{prae} and \ref{hyad}.)

\begin{figure}[!t]
\centerline{\includegraphics[width=\columnwidth]{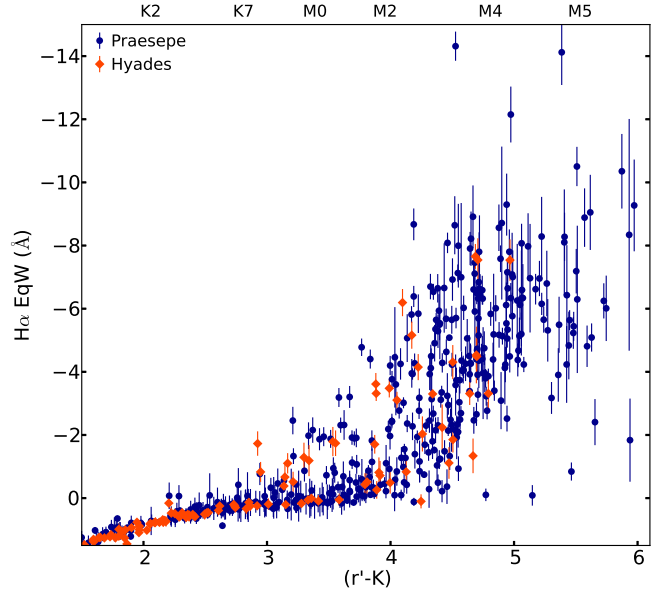}}
\caption{EqW vs.~$(r'-K)$ for stars in Praesepe (blue dots) and the Hyades (orange diamonds). For stars with multiple measurements, the average is shown.  We do not show higher-mass stars with $(r'-K)<1.5$, but the trend of consistent levels of $H\alpha$ activity in the two clusters continues to $(r'-K)\approx1$, the bright limit of our observations in the Hyades.  We find no evidence for different levels of activity in the two clusters (see Section~\ref{res}).} 
\label{fig:eqw} 
\end{figure} 

\begin{figure}[!t]
\centerline{\includegraphics[width=\columnwidth]{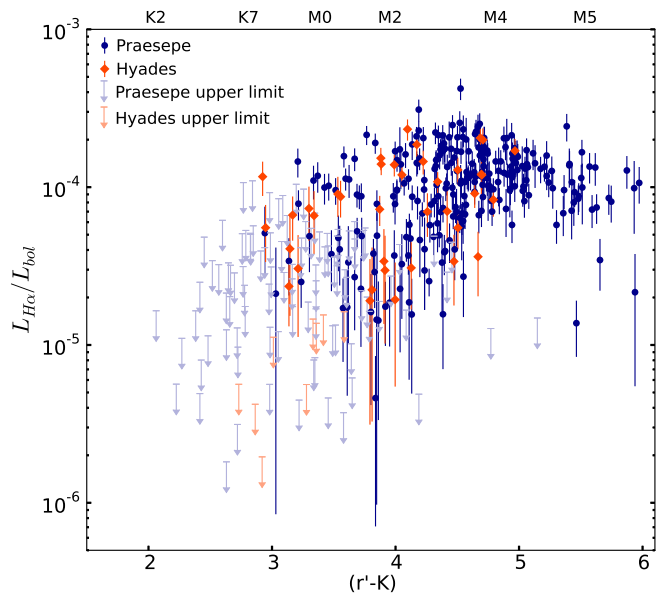}}
\caption{$L_{H\alpha}/L_{bol}$ vs.~$(r'-K)$ for stars in Praesepe and the Hyades, including upper limits. For stars with multiple measurements, the average $L_{H\alpha}/L_{bol}$ is shown. There are no stars with $(r'-K)\ \lapprox$ 3 with definitive $H\alpha$ emission, and all the stars with $(r'-K)\ \lapprox$ 2 are statistically inconsistent with emission at the $2\sigma$ level. The upper envelope of $L_{H\alpha}/L_{bol}$ increases to $(r'-K) \approx 4.5$ before decreasing again; at this color the amount of scatter in $L_{H\alpha}/L_{bol}$ begins to decrease significantly. Our EqW uncertainties are generally smaller for Hyads than for Praesepe stars, placing more stringent upper limits on emission from Hyades stars.} 
\label{fig:lhalbol} 
\end{figure}

Figure~\ref{fig:lhalbol} shows the average $L_{H\alpha}/L_{bol}$ (along with upper limits) as a function of $(r'-K)$ for $P_{mem}\ge70\%$ stars. The scatter in $L_{H\alpha}/L_{bol}$ lessens significantly for $(r'-K)\ \gapprox\ 4.5$. The upper envelope of activity appears to increase slightly with color, peaking at $(r'-K)\approx4$ before decreasing slightly again at the reddest colors.  

\subsection{Stellar Masses}\label{masses}
We estimated masses for every star in our sample using the mass-absolute $K$ magnitude ($M_K$) relation assembled by \citet{adam2007}, who provided masses and spectral energy distributions (SEDs) for B8-L0 stars. We chose this method over the mixed empirical and model-based method used in Paper~I because \citet{adam2007} accounted for observations that models under-predict masses for stars $<$0.5~\Msun. This also had the advantage of giving us a single source for mass calculations across our entire sample. 

For Praesepe, we calculated $M_K$ using a {\it Hipparcos}-derived cluster distance of 181.5$\pm$6.0 pc \citep{vanleeuwen2009}. For the Hyades, we used {\it Hipparcos} parallaxes \citep{hip} where possible to determine distances to individual stars. When {\it Hipparcos} parallaxes were not available, we used the secular parallaxes published by \citet{roser2011}.  The 13 Kundert et al.\ stars that are not in the \citet{roser2011} catalog do not have {\it Hipparcos} parallaxes, and for these stars we assumed a distance of 47 pc \citep{vanleeuwen2009}.  

We determined each star's mass by linearly interpolating between the $M_K$ and mass points given by \citet{adam2007}. The resulting Praesepe masses used in this paper differ by 0.02$-$0.07~\Msun\ from those listed in Paper~I. Masses for all stars in our sample are given in Tables~\ref{prae} and \ref{hyad}. Figure~\ref{fig:periodmass} shows the combined mass-period data for Praesepe and the Hyades, along with the typical mass uncertainties that result from the distance and photometric uncertainties in Praesepe. 

\begin{figure}[t]
\centerline{\includegraphics[width=\columnwidth]{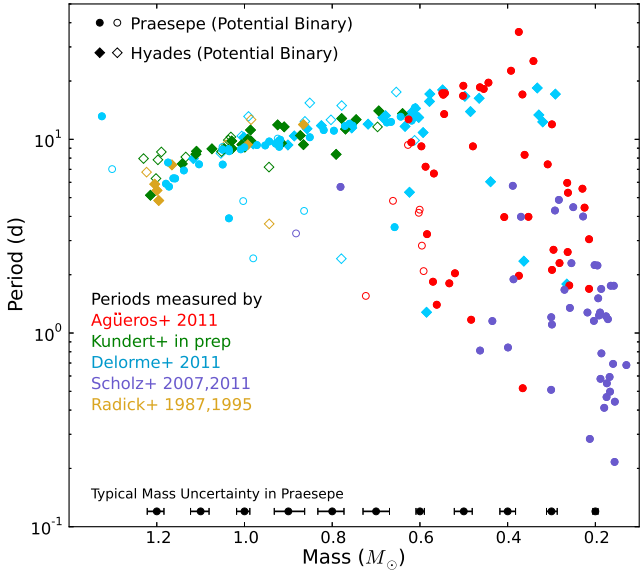}}
\caption{Mass-period diagram for Praesepe and the Hyades. Confirmed binaries from the literature are not shown.  The black points with error bars show the typical mass uncertainties that result from the distance uncertainty and photometric uncertainties for Praesepe members.  All but three stars with $M\ \gapprox\ 0.7~\Msun$ that have not joined their fellow cluster members on the slow-rotator sequence are photometrically identified potential binaries. The three exceptions may be binaries with smaller mass ratios, or they may host giant planets in tight orbits \citep{katja2014,kovacs2014}.}
\label{fig:periodmass} 
\end{figure} 

\begin{figure}[!h]
\centerline{\includegraphics[width=\columnwidth]{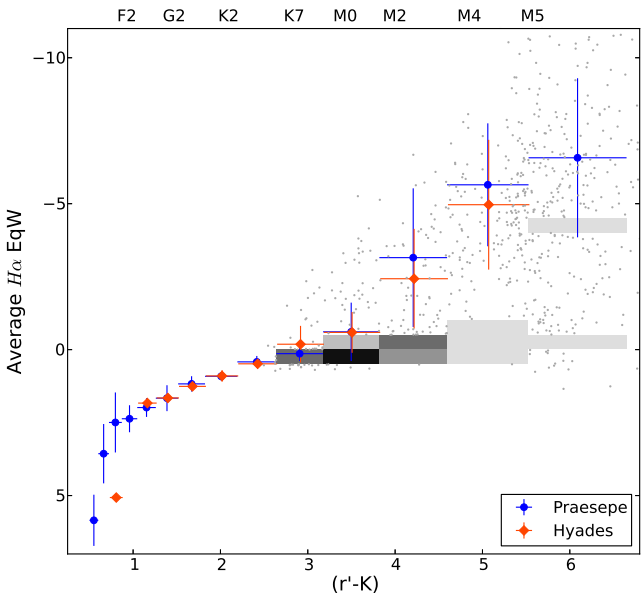}}
\caption{Average \halpha EqW vs.\ logarithmically binned color for stars in Praesepe and the Hyades. The vertical bars show the standard deviation within the bin, and the horizontal bars show the extent of the bin. The $\approx$2800 SDSS M dwarfs \citep[][]{covey2007,west2011} are shown as a grayscale histogram when more than 25 stars fell into a bin, and as gray points otherwise. The inactive region of the histogram includes 2059 stars. KS tests indicate that, for $(r'-K)\ \gapprox\ 2$, the EqWs for Praesepe and the Hyades are consistent with coming from the same distribution, and are inconsistent with the distribution for the field-star sample.} 
\label{fig:field} 
\end{figure} 

\subsection{Binary Identification} \label{binaries}
In Paper~I, we followed \citet{steele1995} in identifying a binary main sequence in the Praesepe CMD offset by 0.75~mag for a given color from that of single stars. We then labeled as candidate binary systems stars that lie above the midpoint between the single-star and binary main sequences \citep{hodgkin1999}.

We applied the same method to both of our CMDs here, but only for stars with $(r'-K)<4$. To the red of this value, the binary main sequence is no longer as apparent, so determining candidate binaries requires more information than a single color and magnitude.  We identified 15 and 29 potential binary systems among known rotators in Praesepe and the Hyades, respectively; these stars are flagged in Tables~\ref{prae} and \ref{hyad}. Four of the possible binary systems in Praesepe were similarly flagged in Paper~I.  Since no color cut was imposed on potential binaries in Paper~I, there were 14 stars flagged as potential binary systems in Paper~I which we did not flag here. Radial velocity monitoring is required to confirm that these are actually binaries. For now, these stars are shown as open symbols in Figure \ref{fig:periodmass} and we removed them when appropriate for our analysis.

Finally, we searched the literature for any confirmed binaries amongst stars with measured $P_{rot}$. We did not find any known binaries in Praesepe. Eight Hyades members were identified in SIMBAD as spectroscopic binaries or as having a M dwarf companion. \citet{delorme2011} also listed an additional spectroscopic binary.  We removed these nine stars from our sample for our analysis.

\subsection{Rossby Numbers} \label{rossby}
Stellar activity evolves with rotation in a mass-dependent way. For stars of a given mass, those rotating above a threshold velocity show emission independent of rotation rate, while below this saturation velocity stars show decreasing activity with decreasing rotation \citep{noyes1984}.  The saturation velocity depends on stellar mass \citep{pizzolato2003}.  Analysis of activity as a function of Rossby number, $R_o =\ P_{rot}/\tau$, where $\tau$ is the convective overturn time, removes this mass-dependence of the rotation-activity relation. 

To calculate \Ro for stars in our sample, we used the equation of \citet{wright2011} for $\tau$ as a function of mass. These authors calculated $\tau$ such that the turnover point for $L_{X}/L_{bol}$ occurs at the same \Ro regardless of stellar mass.  This produces an empirical scaling factor that removes the mass-dependence of the turnover point; we note that this is different from obtaining $\tau$ from comparisons to models.  Tables~\ref{prae} and \ref{hyad} include \Ro values for rotators in the two clusters. 

\section{Results and Discussion}\label{res}
\subsection{Comparing Chromospheric Activity in the Two Clusters} \label{similar}
The data in Figure \ref{fig:eqw} indicate that Praesepe and the Hyades have similar levels of chromospheric activity.  Stars with $2<(r'-K)<3$ do not have statistically significant levels of \halpha emission; some stars in this range have \halpha EqWs consistent with emission, but many of those are potential binaries.  Emission is more reliably detected starting at $(r'-K)\approx3$, or SpTs of $\approx$K7.  All stars with $(r'-K)\ \gapprox\ 4.5$ (later than $\approx$M3) appear to be active, and the two clusters visually appear to have similar upper and lower envelopes of activity.  

Figure~\ref{fig:lhalbol} is a comparison between the $L_{H\alpha}/L_{bol}$ for both clusters; the clusters also appear to have consistent levels of activity by this measure. The upper limits in Figure~\ref{fig:lhalbol} are slightly misleading because our Hyades stars have smaller EqW errors, likely because stars of the same mass have apparent magnitudes $\approx$3 mag brighter in the Hyades than in Praesepe (see Figure~\ref{fig:psource}). The correspondingly higher S/N for those spectra allows us to place more stringent upper limits on \halpha emission in the Hyades than in Praesepe.

Figure~\ref{fig:field} shows the average EqW for each cluster as a function of binned $(r'-K)$. It also includes EqWs for nearly 2800 SDSS M dwarfs; we constructed this sample by cross-matching the \citet{west2011} M-dwarf catalog with the ``high quality'' sample of SDSS/2MASS photometry from \citet{covey2007}.\footnote{This sample includes two Praesepe stars.} We use logarithmic bins in $(r'-K)$ because we have more high-mass stars than low-mass stars in the Hyades; the bins increase in size for redder colors but still contain approximately the same number of stars (between eight and 20).  

Kolmogorov-Smirnov (KS) tests in each color bin find that for $(r'-K)\ > 1$, the EqWs for the cluster stars are consistent with coming from the same distribution. Furthermore, for $(r'-K)\ > 2.6$, these EqWs are inconsistent with the distribution of EqWs for the low-activity (and on average, older) field-stars. (The exception is the $2.6<(r'-K)<3.2$ bin, where the Hyads are consistent with the field stars (p $ =$ 0.14).)  It is therefore appropriate to treat the two clusters as a single-aged cluster for purposes of analysis, as we do below. 

Figure~\ref{fig:field} also shows clearly that the late-type cluster stars are systematically more active than their SDSS counterparts. The field star ages are not known, but they presumably range between 2$-$10 Gyr.  These data therefore illustrate nicely the overall decay of magnetic activity with time \citep[as noted by e.g.,][]{skumanich72,radick1987, soderblom2001}.

\begin{figure}[t!]
\centerline{\includegraphics[width=\columnwidth]{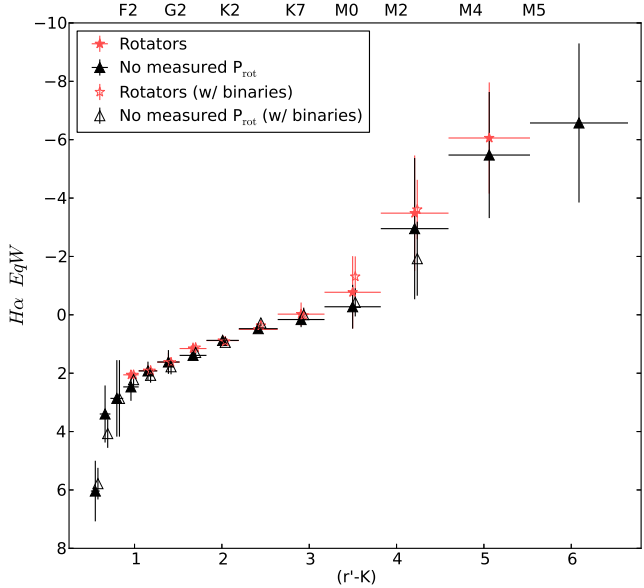}}
\caption{\halpha EqWs for stars with and without measured $P_{rot}$, using the same logarithmic color bins as in Figure \ref{fig:field}. The data with potential binaries included (open symbols) have been offset slightly for clarity. KDE tests show that our sample of rotators is not biased toward more active stars: the distribution of \halpha EqWs is similar regardless of whether the stars have a measured $P_{rot}$.}
\label{fig:comprot} 
\end{figure} 

How do our results compare to previous authors' comparisons of chromospheric activity in Praesepe and the Hyades? \citet{pace2004} found that solar-type stars in the two clusters have similar levels of chromospheric activity, as measured by CaII K emission. Our data are consistent with this result, and extend it to later-type stars. 

However, \citet{kafka2006} found that \halpha activity in the Hyades began at bluer colors than in Praesepe. \citet{kafka2006} also found that the Hyads in their sample became completely active at a bluer color than those in Praesepe. Because these authors combined their \halpha measurements in Praesepe with literature EqWs for both clusters, it is possible that the disagreement is due to inconsistencies in the methods used to measure EqWs. As discussed in Section \ref{activity}, our \halpha EqWs are systematically 0.1$-$1~\AA~larger than those measured by \citet{kafka2006} for the same stars in Praesepe. Shifting the upper envelope of Praesepe EqWs up by $\approx$0.5~$\AA$~in figure 7 of \citet{kafka2006} would essentially remove the difference in the location of the transition between inactive and active stars in the two clusters reported by these authors. Such a shift, however, would not change the color at which all of the Praesepe stars become active.

\begin{figure}[!b]
\centerline{\includegraphics[width=\columnwidth]{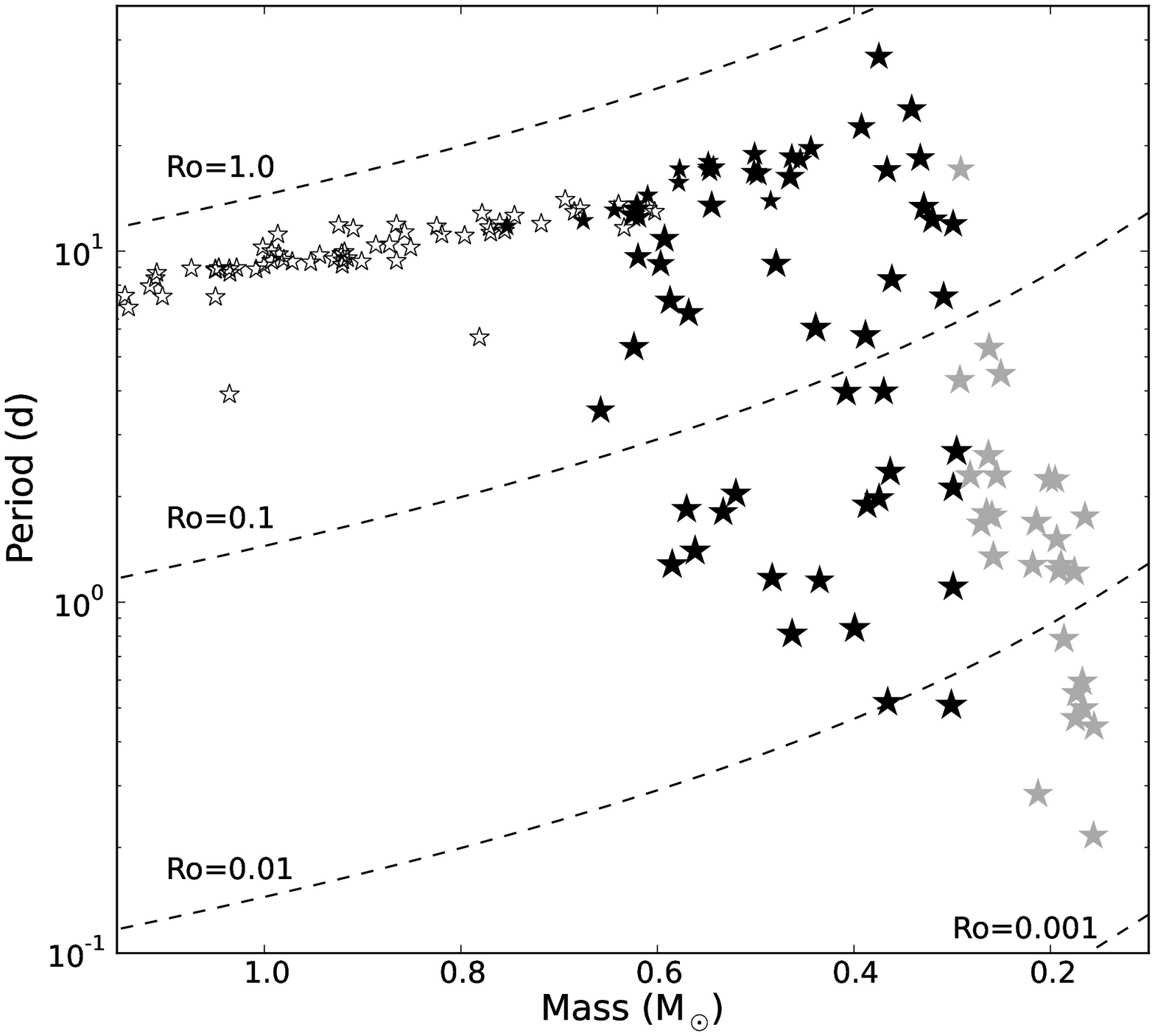}}
\caption{Mass-period diagram for Praesepe and Hyades stars with measured $P_{rot}$ and \halpha EqW and $P_{mem}\ge70\%$ (known and candidate binaries are not included). Fully convective stars ($\leq$0.3~\Msun) are in gray. Solid symbols indicate stars with \halpha in emission; nearly all stars with $M\ \lapprox\ 0.6$~\Msun\ are active. Lines of constant \Ro are plotted for reference. Only one fully convective star has \Ro $> 0.1$; given the uncertainties in the masses, it may well have $M > 0.3$~\Msun.}
\label{fig:rossby} 
\end{figure} 

\subsection{Activity and Measurements of Periodic Variability}\label{bias}
Active stars may have higher spot coverage and might therefore show stronger photometric modulation than non-active stars, which would bias our sample of rotators toward stars with stronger $H\alpha$ emission.  If \halpha active stars are more likely to exhibit periods, then the age-rotation-activity relation derived from stars with measured periods (rotators) may not apply to stars without measured periods.

On the other hand, periodic variability may not be detected for a variety of reasons. Stars without measured $P_{rot}$ may have photometric variability that falls below the detection threshold of a given survey, which in turn may be due to a lack of spots or to symmetrical spot coverage across the stellar surface. They may also have $P_{rot}$ that is too short or too long to be detected by that survey.

Figure~\ref{fig:comprot} shows a comparison of the \halpha EqW as a function of $(r'-K)$ for stars in our sample with and without detected $P_{rot}$. To test the similarity between these two samples, we determine the probability that the stars with detected $P_{rot}$ are drawn from the same distribution as the stars without detected $P_{rot}$ using the procedure outlined in \citet{cargile2014}. To begin, we derived the probability distribution function (PDF) for both EqW-color distributions using the Gaussian kernel density estimation (KDE) function from the SciPy package.\footnote{scipy.stats.gaussian\_kde, \url{http://www.scipy.org/}}  Instead of binning the data, KDE uses a kernel function to smooth over all the data points and produce a continuous distribution \citep[see ][ for details]{silverman1986}. We used an automatic bivariate bandwidth determination based on ``Scott's rule'' \citep{scott1992} to choose the kernel width.  

Once we had the two PDFs, we multiplied them together and integrated the product over the full parameter space.  This gave a metric that describes the overall correlation between the distributions of stars with and without detected $P_{rot}$.  If potential binaries are included in the test, $p_{0,all}=0.0420$, but $p_{0,no\ bin}=0.03798$ if potential binaries are removed.

To find the significance of this metric, we used Monte Carlo simulations.  First, we randomly drew $10^5$ sub-samples of 194 stars without $P_{rot}$ measurements (the same number as our sample of rotators) and compared this to the full distribution of stars without $P_{rot}$. This showed what metric results if the rotators are actually drawn from the same distribution as the stars without $P_{rot}$ measurements.  Then we randomly drew $10^5$ sub-samples of 194 stars from a flat distribution over the observed space and calculated the average probability metric again.  This showed what metric results if the rotators come from a random distribution.  The average probability metric in the first case is $p_{1,all}=0.0365$, and in the second case it is $p_{2,all}=0.0067$.  If potential binaries are excluded, we find $p_{1,no\ bin}=0.0336$ and $p_{2,no\ bin}=0.0066$  Thus, we concluded that the rotator distribution is more likely to be drawn from the non-rotator distribution than from a random distribution in the same observed space.}

This implies that our sample of rotators is not biased toward stars with stronger \halpha activity. We can therefore use our sample of rotators for which we have measured \halpha EqWs to characterize the relationship between activity and rotation for all 600~Myr stars, regardless of whether they exhibit periodic behavior at any given epoch.  

\subsection{The Relationship Between $H\alpha$ Emission and Rotation} \label{halpha}
We have assembled a large sample of stars from Praesepe and the Hyades to test the rotation-activity relation at 600 Myr.  As Figure \ref{fig:periodmass} shows, the only notable difference between the two clusters' period-mass distributions is that there are no known $<$0.26~\Msun\ rotators in the Hyades, and only two $\leq$0.3~\Msun. By contrast, our lowest-mass Praesepe rotator has $M=0.15$~\Msun, and we have spectra for 28 Praesepe stars with $0.15\le M\le0.3$~\Msun. We therefore are dependent mostly on Praesepe stars for any analysis of activity and rotation in fully convective stars at this age.  However, since the distributions of activity versus color are consistent between the two clusters (Section \ref{similar}), and our sample of rotators is not biased toward more active stars (Section \ref{bias}), we can use the combined sample as a proxy for all 600 Myr stars. In Figure~\ref{fig:rossby}, we reproduce the mass-period diagram for both clusters and highlight \halpha active and fully convective ($M < 0.3$~\Msun) stars.

\begin{figure}[t!]
\centerline{\includegraphics[width=\columnwidth]{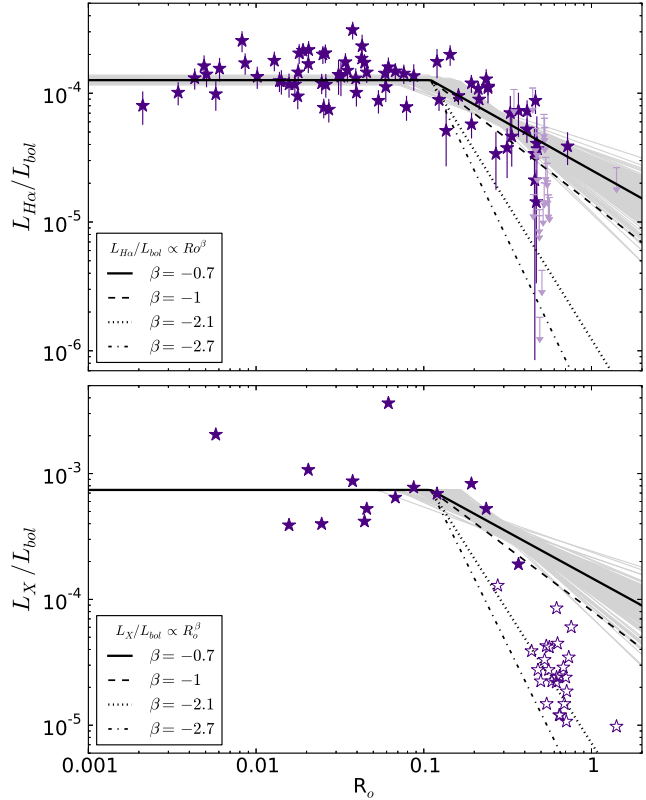}}
\caption{{\it Top ---} $L_{H\alpha}/L_{bol}$ vs.\ $R_o$ for Praesepe and Hyades stars with various power-laws for the unsaturated regime overlaid (for \Ro $\gtrsim 0.11$).  The results from the fit discussed in the text are also shown: the gray region shows 200 models drawn randomly from the posterior probability distribution, and the solid black line is the maximum {\it a posteriori} model. Upper limits are shown but not included in the fit. 
The broken black lines show power-laws from the literature.  {\it Bottom} --- $L_X/L_{bol}$ vs.\ $R_o$ for Praesepe and Hyades stars \citep[data from][]{wright2011}. Empty symbols indicate stars $>$0.68~\Msun, the highest mass at which \halpha emission is detected. The \halpha data are consistent with a shallow decline of activity with rotation ($\propto$$R_o^{-1}$), while X-ray activity appears to decline in a manner more consistent with the steeper \citet{randich2000b} or \citet{wright2011} power-laws ($\propto$$R_o^{-2.1}$ and $R_o^{-2.7}$, respectively).}
\label{fig:slopes} 
\end{figure}

The top panel of Figure~\ref{fig:slopes} shows $L_{H\alpha}/L_{bol}$ as a function of \Ro for all observed rotators with $P_{mem}\ge70\%$; it includes $2\sigma$ upper limits for stars whose \halpha EqW is consistent with absorption. Stars with $M > 0.3~\Msun$ appear to follow a saturation-type rotation-activity relation: for $\Ro\ \lapprox\ 0.11$, the activity is approximately constant. This result is consistent with prior results that stars from mid-F to early-M SpTs exhibit a saturation-type relationship between rotation and chromospheric activity \citep{noyes1984,delfosse1998,jackson2010}. 

At larger $R_o$, activity decreases with increasing $P_{rot}$ and increasing $R_o$. However, our data hint that this may not be a smooth power-law decline: the slowly rotating stars with \Ro $\gapprox\ 0.45$ suggest a sharp decrease in chromospheric activity over a small range in $R_o$. Because we can only give upper limits on $L_{H\alpha}/L_{bol}$ for most of these stars, our data do not allow us to confidently claim this change in behavior, and further activity measurements are required to investigate activity for slow rotators.

Nearly all the fully convective stars in our sample have \Ro $\le 0.07$ and saturated levels of \halpha activity. One has \Ro $> 0.2$; it is the slowest rotator among the $<$0.3~\Msun\ stars in Figure~\ref{fig:rossby}. This outlier has $M=0.291$ \Msun. However, given the uncertainties in the masses, it may well have $M > 0.3$~\Msun, and its rotation-activity behavior is consistent with that of the higher-mass stars. Aside from this outlier, all the stars with $M\le0.3~\Msun$ are rotating fast enough to have saturated levels of activity.  

We therefore parametrize the rotation-activity relationship for our stars as a flat region connected to a power-law.  Below the turnover point ($\Ro_{,sat}$), activity is constant and equal to $(L_{H\alpha}/L_{bol})_{sat}$. Above $\Ro_{,sat}$, activity declines as a power-law with index $\beta$.  Functionally, this corresponds to
\begin{equation*}
\frac{L_{H\alpha}}{L_{bol}} = 
\begin{cases}
\left( \frac{L_{H\alpha}}{L_{bol}} \right)_{sat}, & \text{if } \Ro\leq \Ro_{,sat} \\
C \Ro^{\beta}, & \text{if } \Ro>\Ro_{,sat}
\end{cases}
\end{equation*}
where $C$ is a constant. This model has been widely used in the literature \citep[e.g.,][]{randich2000b, wright2011}.

\begin{figure}[t!]
\centerline{\includegraphics[width=\columnwidth]{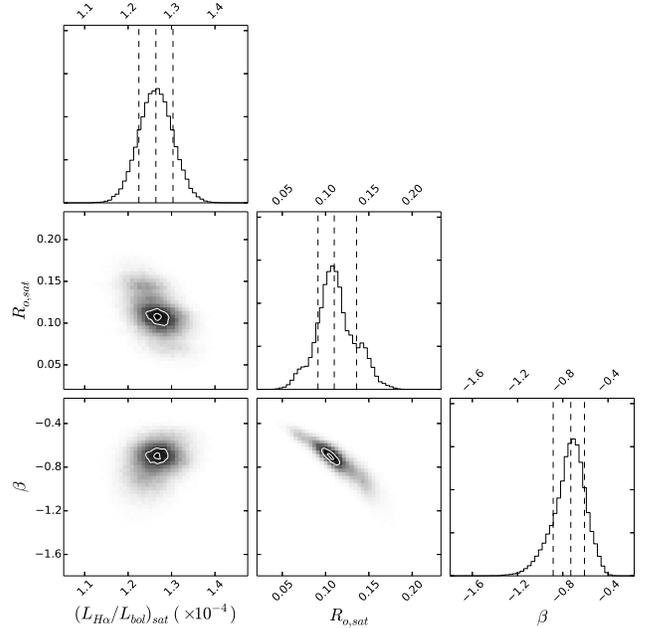}}
\caption{The marginalized posterior probability distributions from the MCMC analysis using {\it emcee}.  The peaks of the 1D distributions correspond to the maximum {\it a posteriori} model, and the two-dimensional (2D) distributions illustrate covariances between parameters. The vertical lines on the 1D histograms indicate the median and 68-percentile values; the contours on the 2D histograms indicate the 99.5- and 68-percentile of the distributions.}
\label{fig:corner} 
\end{figure}

We used the open-source Markov-chain Monte Carlo (MCMC) package {\it emcee} \citep{dfm2013} to fit the three-parameter model described above to our data.  The fit derives posterior probability distributions over each parameter; these distributions are shown in Figure \ref{fig:corner}. Figure \ref{fig:slopes} includes 200 random models drawn from these distributions; each gray line represents a model that fits the data, though it is not the most probable model. Figure \ref{fig:slopes} also shows the maximum {\it a posteriori} model, which is the most probable model.

The parameters corresponding to the maximum {\it a posteriori} model are $(L_{H\alpha}/L_{bol})_{sat}=(1.26\pm0.04) \times 10^{-4}$, $\Ro_{,sat}=0.11^{+0.02}_{-0.03}$, and $\beta=-0.73^{+0.16}_{-0.12}$, where the stated values correspond to the 50$^{\rm th}$ quantile of the results and the uncertainties correspond to the 16$^{\rm th}$ and 84$^{\rm th}$ quantiles, respectively.  We selected these quantiles to be consistent with 1$\sigma$ Gaussian uncertainties, even though our one-dimensional 1D posterior probability distributions are not Gaussian. We also note that $\Ro_{,sat}$ and $\beta$ are highly anti-correlated: a lower $\Ro_{,sat}$ results in a shallower $\beta$, and vice versa. We find a turnover point $\Ro_{,sat}$ consistent with that found in the literature \citep[e.g.,][]{wright2011}, but the $\beta$ we derive is inconsistent with literature values \citep[e.g.,][]{jackson2010} by 2$-$11$\sigma$.  

\subsection{Chromospheric and Coronal Activity-Rotation Relations at 600 Myr}
A number of authors have also derived power-laws to describe the unsaturated rotation-activity regime for other activity indices. \citet{jackson2010} observed a saturation-type relationship between chromospheric emission (measured using CaII) and rotation in early M dwarfs in the younger, $\approx$150 Myr-old cluster NGC~2516. Their figure 9 shows saturated activity below $\Ro_{,sat}\approx0.1$; above that, activity decreases as a power-law with $\beta\approx-1$.  In her summary of $\ROSAT$ results for open clusters and field stars, \citet{randich2000b} found that the data for \Ro $\gapprox\ 0.16$ were best fit by a $\beta = -2.1\pm0.09$ power-law. More recently, \citet{wright2011} found that, for \Ro $> 0.13$, the decline in coronal activity followed a slightly steeper $\beta=-2.18\pm0.16$ power-law; furthermore, these authors calculated $\beta=-2.70\pm0.13$ for a set of solar-type stars.  

Our \halpha data are best fit by a power-law that is clearly shallower than and inconsistent with the three power-laws described above, as shown in the top panel of Figure~\ref{fig:slopes}.  The \citet{jackson2010} value of $\beta=-1$, which also describes chromospheric activity for young, low-mass stars, comes closest to describing our data. Our data, however, are inconsistent with this value at the 2$\sigma$ level, and are better fit by an even shallower power-law.  

The bottom panel of Figure~\ref{fig:slopes} shows $L_X/L_{bol}$, calculated using the X-ray data for Praesepe and Hyades stars published by \citet{wright2011}, as a function of $R_o$. The shallower power-laws we find best describe our $L_{H\alpha}/L_{bol}$ data are not consistent with the behavior of $L_X/L_{bol}$ for most X-ray-emitting stars in these clusters, which appear to follow a steeper power-law relation: our $L_{H\alpha}/L_{bol}$ data are inconsistent with the \citet{randich2000} and \citet{wright2011} relationships for unsaturated stars at the 7$-$11$\sigma$ level.

This is not entirely surprising: \citet{preibisch2005} and \citet{stelzer2013} found that X-ray emission declines more rapidly than chromospheric activity indicators with age, both for solar-type stars and M dwarfs. In our sample, one possible explanation for the difference in the unsaturated behavior of $L_{H\alpha}/L_{bol}$ and $L_X/L_{bol}$ for stars is that the subsets with \halpha and X-ray detections have different mass distributions. The unsaturated $H\alpha$-emitters in the top panel of Figure~\ref{fig:slopes} range from $\approx$0.4$-$0.7~\Msun, while the unsaturated X-ray-emitting stars in the bottom panel are mostly $\approx$1~\Msun.

The few X-ray-emitting stars in the bottom panel of Figure~\ref{fig:slopes} that have $M<0.7$ \Msun\ and $\Ro>\Ro_{,sat}$ also suggest a mass-dependent rotation-activity relationship; these stars lie closer to the shallow power-laws derived for chromospheric emission from low-mass stars. Although using \Ro should provide a mass-independent way to examine rotation and activity \citep{pizzolato2003,wright2011}, it may not remove this dependence entirely: unsaturated emission may decline with increasing $R_o$ at different rates for stars of different masses.

As noted by \citet{covey2008}, a separate problem with these comparisons is that these are non-simultaneous \halpha and X-ray measurements. The \citet{wright2011} sample may be preferentially selecting stars in an X-ray flare state, while the \halpha measurements may have been taken when the star has returned to a quiescent state. While our data hint at underlying mass- and age-related differences in the evolution of chromospheric and coronal emission, our sample is small. A larger sample of X-ray measurements in Praesepe and the Hyades, ideally made simultaneously with \halpha measurements, is required to draw firmer conclusions.

\section{Conclusion}\label{concl}
\begin{enumerate}
\item We have collected 720 spectra of 516 high-confidence Praesepe members, and 139 spectra of 130 high-confidence Hyads; more than half of the Praesepe spectra and all of the Hyades spectra are new observations. We have measured \halpha EqWs for all of these spectra, estimating the EqW uncertainties by accounting for both human measurement error and for photon noise. 
\item To convert these \halpha EqWs into mass-independent $L_{H\alpha}/L_{bol}$ values, we have computed our own $\chi$ factors. $\chi = f_0/f_{bol}$, where $f_0$ is the continuum flux level for the $H\alpha$ line and $f_{bol}$ is the apparent bolometric flux. Our values differ from those presented in \citet{walkowicz2004} and \citet{west2008}; see Appendix~\ref{appen} for details. 
\item We have found that Praesepe and the Hyades follow a nearly identical color-activity relation, implying that they have very similar ages. This contradicts the results of  \citet{kafka2006}, who found that activity in the Hyades began at bluer colors than in Praesepe, and that Hyads became completely active at a bluer color than stars in Praesepe. Because \citet{kafka2006} combined their \halpha measurements in Praesepe with literature EqWs for both clusters, it is possible that the disagreement is due to inconsistencies in the methods used to measure EqWs. Our results are consistent with the finding of \citet{pace2004} that  solar-type stars in the two clusters have similar levels of chromospheric activity, as measured by CaII K emission.
\item We gathered $P_{rot}$ for 135 Praesepe members and 87 Hyads from PTF observations and from the literature. Taking the two clusters as a single-aged sample, we constructed a combined mass-period distribution for stars at 600 Myr. 
We examined the \halpha EqWs of known rotators and of stars without a measured $P_{rot}$ in our sample, finding that the known rotators are not more active, on average, than the stars without measured periods. We can therefore use our sample of rotators for which we have measured \halpha EqWs to characterize the relationship between activity and rotation for all 600~Myr stars.
\item We have demonstrated the presence of a \Ro$\approx0.11$ chromospheric activity threshold for low-mass stars at 600 Myr. Stars rotating below this threshold show saturated levels of activity, and stars with slower rotation speeds show declining activity levels.
\item We have presented preliminary evidence that chromospheric activity (as measured by \halpha) and coronal activity (measured by X-ray emission) decline differently as a function of $R_o$.
\end{enumerate}

\citet{jackson2010} found that, at 150 Myr, fully convective M dwarfs showed CaII emission at levels roughly independent of rotation. These observations of saturated, fully convective M dwarfs at 150 Myr, and now at 600 Myr, differ somewhat from what is seen in the field. Some field M dwarfs with $M\ \lapprox\ 0.3$~\Msun\ have longer $P_{rot}$ and follow an unsaturated rotation-activity relationship for \Ro $\gtrsim 0.1$ \citep{mohantybasri2003,wright2011}. Low-mass members of Praesepe and the Hyades are young enough to rotate faster than the saturation velocity for fully convective M dwarfs. As they age, these stars should begin to spin down into the unsaturated regime observed for field stars. Similar studies of older clusters are essential to map out fully the evolution of the chromospheric activity-rotation relation for these low-mass stars. 

However, while a dozen open clusters with ages $\lapprox$600~Myr have been extensively surveyed both for tracers of magnetic activity and for rotation, few clusters older than the Hyades and Praesepe and younger than field stars, whose ages are imprecisely known but range from 2-10 Gyr, have been studied in the same detail. Recent work on NGC~752 (Ag\"ueros et al., in prep) and on the three open clusters in the {\it Kepler} field of view \citep[including NGC 6811;][]{meibom2011} will add to our knowledge of stellar properties at $\gtrsim$1 Gyr. For now, the clusters at $\approx$600 Myr anchor the transition from young open clusters to more rare evolved clusters and field stars. The results of our examination of activity and rotation in the Hyades and Praesepe are therefore an essential data point in the study of the evolution of these properties.  

\acknowledgments We thank the anonymous referee for comments that improved the paper. We thank John Thorstensen for his help with the MDM observations and the WIYN observing specialists for their assistance. We thank Stella Kafka and John Stauffer for sharing their spectra with us.  We thank Sarah Schmidt, Keivan Stassun, Lucianne Walkowicz, Andrew West, and Nicholas Wright for useful discussions and comments.  M.A.A.\ acknowledges support provided by the NSF through grant AST-1255419.  P.A.C.\ acknowledges support provided by the NSF through grant AST-1109612.

This research has made use of NASA's Astrophysics Data System Bibliographic Services, the SIMBAD database, operated at CDS, Strasbourg, France, the NASA/IPAC Extragalactic Database, operated by the Jet Propulsion Laboratory, California Institute of Technology, under contract with the National Aeronautics and Space Administration, and the VizieR database of astronomical catalogs \citep{Ochsenbein2000}. 

The Two Micron All Sky Survey was a joint project of the University of Massachusetts and the Infrared Processing and Analysis Center (California Institute of Technology). The University of Massachusetts was responsible for the overall management of the project, the observing facilities and the data acquisition. The Infrared Processing and Analysis Center was responsible for data processing, data distribution and data archiving.  

Funding for SDSS-III has been provided by the Alfred P. Sloan Foundation, the Participating Institutions, the National Science Foundation, and the U.S. Department of Energy Office of Science. The SDSS-III web site is http://www.sdss3.org/.

SDSS-III is managed by the Astrophysical Research Consortium for the Participating Institutions of the SDSS-III Collaboration including the University of Arizona, the Brazilian Participation Group, Brookhaven National Laboratory, University of Cambridge, Carnegie Mellon University, University of Florida, the French Participation Group, the German Participation Group, Harvard University, the Instituto de Astrofisica de Canarias, the Michigan State/Notre Dame/JINA Participation Group, Johns Hopkins University, Lawrence Berkeley National Laboratory, Max Planck Institute for Astrophysics, Max Planck Institute for Extraterrestrial Physics, New Mexico State University, New York University, Ohio State University, Pennsylvania State University, University of Portsmouth, Princeton University, the Spanish Participation Group, University of Tokyo, University of Utah, Vanderbilt University, University of Virginia, University of Washington, and Yale University.

\appendix
\section{Recalculating the $\chi$ Factor}\label{appen}
\citet{walkowicz2004}, hereafter WHW04, describe the difficulty of computing $L_{H\alpha}/L_{bol}$: it requires flux-calibrated spectra and measurements of each star's distance.  WHW04 therefore derive a distance-independent method for calculating $L_{H\alpha}/L_{bol}$ using the factor $\chi = f_0/f_{bol}$, where $f_0$ is the continuum flux level for the $H\alpha$ line and $f_{bol}$ is the apparent bolometric flux. When the $H\alpha$ line EqW, $W_{H\alpha}$, is known, one may find
\[
\frac{L_{H\alpha}}{L_{bol}} = -W_{H\alpha} \chi .
\]
Calculating $\chi$ for a star requires photometry and a well-calibrated spectrum, which provide the means for calculating $f_0$ and $f_{bol}$.  We define $f_0$ to be the mean flux in two windows: 6550$-$6560 and 6570$-$6580~\AA~\citep{west2008}.  We determine $f_{bol}$ by finding the bolometric correction and calculating the apparent bolometric magnitude, $m_{bol}$, then converting $m_{bol}$ to $f_{bol}$ using solar values.  The absolute solar bolometric flux, $F_{\odot,bol}$, is  
\[
F_{\odot,bol} = \frac{\Lsun}{4 \pi (10\ {\rm pc})^2} ,
\]
where $\Lsun=3.842\times10^{33}$ erg s$^{-1}$ cm$^{-2}$ is the luminosity of the Sun \citep[][]{mamajek}.  Given the absolute bolometric magnitude of the Sun, $M_{\odot,bol} = 4.74$ \citep[][]{mamajek}, we can then calculate
\[
f_{bol} = F_{\odot,bol} 10^{-0.4 (m_{bol}-M_{\odot,bol})}
\]
for each star. 

WHW04 calculated $\chi$ for early-mid M dwarfs using stars from the 8 pc sample that have spectra in the PMSU spectroscopic survey \citep[][]{reid1995}.  WHW04 used the \citet{leggett1996} $K$-band bolometric corrections to determine $f_{bol}$.  To calculate $f_0$, they used the region 6555$-$6560~$\AA$. \citet{west2008} use the WHW04 $\chi$ values, but define the continuum windows as 6550$-$6560 and 6570$-$6580 \AA. Additionally, WHW04 state that the spectra have been spectrophotometrically calibrated in the region around $H\alpha$, but \citet{reid1995} state that the PMSU spectra are only calibrated to a relative, not an absolute, flux scale.  

\begin{figure}[t!]
\centerline{\includegraphics[width=\columnwidth]{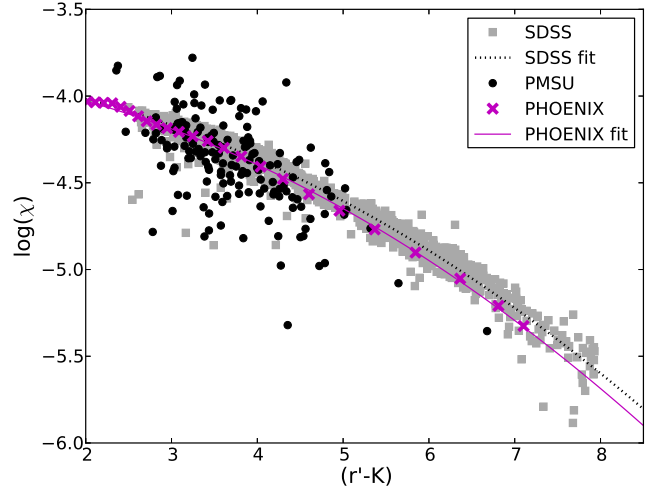}}
\caption{$\chi$ calculated for SDSS and PMSU data and for PHOENIX model spectra.  The PMSU data show significant scatter in $\chi$ and no obvious dependence of $\chi$ on color, while the SDSS data show less scatter and a smooth trend with color. The dotted line shows quadratic fit to the SDSS data. The solid line shows a quadratic fit to the PHOENIX model values, which are similar to, though slightly lower than, the typical SDSS values.} 
\label{fig:compchi} 
\end{figure} 

\begin{figure}[t!]
\centerline{\includegraphics[width=\columnwidth]{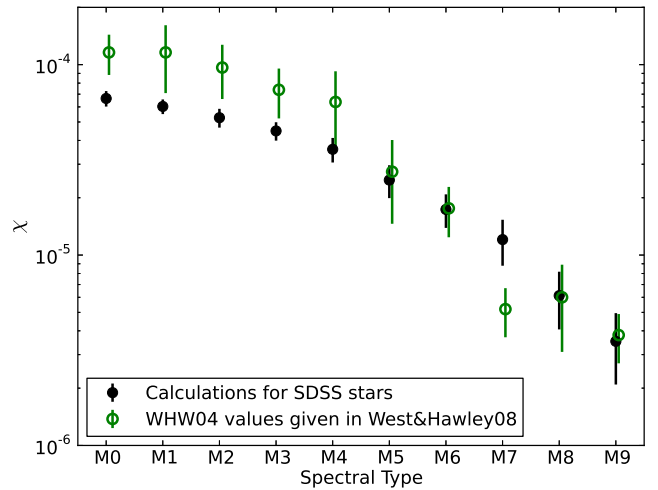}}
\caption{A comparison of our $\chi$ values for SDSS M dwarfs and those listed in \citet{west2008}. The \citet{west2008} values are shifted slightly to the right for clarity. For M0-M3 and M7 dwarfs, our values are inconsistent with those calculated by WHWO4 and \citet{west2008}.}
\label{fig:chi} 
\end{figure}

\begin{figure}[!t]
\centerline{\includegraphics[width=\columnwidth]{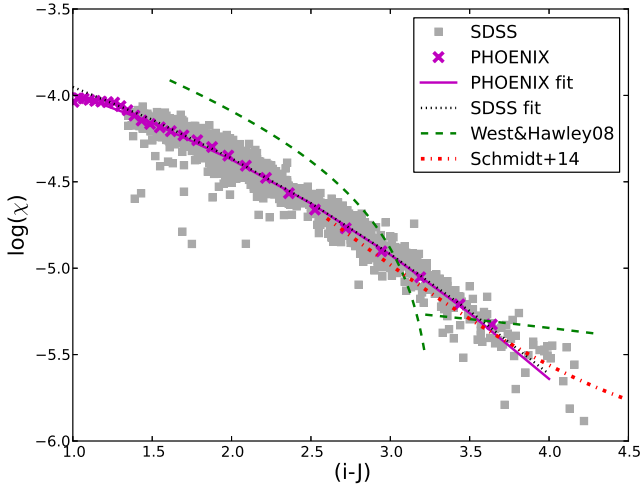}}
\caption{$\chi$ calculated for SDSS M dwarfs and for PHOENIX model spectra. Also shown are the functions for $\chi$ vs. $(i-J)$ given by \citet{west2008} and \citet{schmidt2014}, and quadratic fits to the SDSS data and the PHOENIX model values. The PHOENIX values match the SDSS values very well, but the function from \citet{west2008} shows a very different trend of $\chi$ with color.} 
\label{fig:compimJ} 
\end{figure} 

\begin{deluxetable}{cc}[!t]
\tabletypesize{\scriptsize}
\tablecaption{$\chi$ values for M stars, calculated from SDSS data \label{sdsschi}}
\tablehead{
\colhead{SpT}  &  \colhead{$\chi\  (\times10^{-5})$}  
}
\startdata
M0  &  $6.6453\pm0.6207$ \\ 
M1  &  $6.0334\pm0.5326$ \\ 
M2  &  $5.2658\pm0.5963$ \\ 
M3  &  $4.4872\pm0.4967$ \\ 
M4  &  $3.5926\pm0.5297$ \\ 
M5  &  $2.4768\pm0.4860$ \\ 
M6  &  $1.7363\pm0.3475$ \\ 
M7  &  $1.2057\pm0.3267$ \\ 
M8  &  $0.6122\pm0.2053$ \\ 
M9  &  $0.3522\pm0.1432$
\enddata
\tablecomments{The $\chi$ values are the average of the SDSS distribution, and uncertainties are the standard deviation of that distribution.}
\end{deluxetable}

We computed $\chi$ for PMSU stars using photometry compiled by I.~N.~Reid\footnote{\url{http://www.stsci.edu/~inr/pmsu.html}} and following the same procedure as WHW04, and we found two problems. First, our $\chi$ values for early M dwarfs do not match those calculated by WHW04: our values were systematically lower by $\approx$1/3 dex. After ruling out input photometry, bolometric corrections, and the continuum window for calculating $f_0$ as potential sources of the offset, we traced the discrepancy to a difference in calculating $f_{bol}$.  Our values for $log(f_{bol})$ were consistently $\approx$0.5 dex lower than those given by WHW04's equation 8, which gives $f_{bol}$ as a function of apparent $K$.\footnote{WHW04 give equations to calculate a star's bolometric flux as a function of its apparent magnitude and as a function of its color. These equations are not self-consistent, as apparent flux is distance-dependent, just like apparent magnitude, but color and absolute flux are distance-independent.} We compared our $f_{bol}$ values to a number of model-derived and empirical values \citep[][in addition to PHOENIX- and SDSS-based values; see discussion below]{baraffe98, adam2007, dotter2008, casagrande2008, boyajian2012} and in all cases our calculations matched the literature values, while the WHW04 results were consistently too low.  

Second, the PMSU stars show a large scatter in $\chi$, and therefore the WHW04 $\chi$ values do not have a well-defined dependence on color. This, along with the fact that we did not have a well-calibrated set of spectra to calculate $\chi$ for K dwarfs, motivated our calculation of $\chi$ from the PHOENIX models. To check the model values against empirical data, we also calculated $\chi$ for M dwarfs in SDSS.  After matching the spectroscopic \citet{west2011} SDSS M dwarf catalog with the ``high quality'' SDSS/2MASS photometry of \citet{covey2007}, we calculated $\chi$ for nearly 2800 stars. (We use $BC_r$, calculated as a function of $(r-K)$ from the SED table in \citet{adam2007}, to obtain $f_{bol}$.) The results from both the PMSU and SDSS samples are shown in Figure \ref{fig:compchi}, where $\chi$ has been calculated by our procedure for all stars in that figure. The $\chi$ data for the SDSS stars show much less scatter than for the PMSU sample, and also a clearer color-dependence.

We also calculated the average $\chi$ for SDSS stars of each M subtype \citep[using the SpTs listed in][]{west2011} and compared these $\chi$ values with those presented in WHW04 and \citet{west2008}.  The average and standard deviation of $\chi$ for each SpT is given in Table~\ref{sdsschi}, and plotted with the WHW04 and \citet{west2008} values in Figure~\ref{fig:chi}. For M0-M3 and M7 dwarfs, our values are inconsistent with those calculated by these authors.

To obtain $\chi$ for the full range of K and M dwarfs in our sample, we calculated $\chi$ from the PHOENIX ACES model spectra \citep{husser2013}. We used spectra with $2500 \le T_{eff}\le 4800$ K, $log(g) = 5.0$, and solar metallicity. The continuum flux is measured between 6550$-$6560 and 6570$-$6580 \AA, as above. We computed synthetic photometry by convolving these model spectra with the SDSS and 2MASS filter curves. The resulting optical and near-infrared colors and the corresponding $\chi$ values are given in Table~\ref{phoenix}.  

We fit a quadratic function to $\chi$ versus $(r'-K)$ and $\chi$ versus $(i-J)$, assuming 10\% errors in $\chi$ and using typical photometric errors for SDSS and 2MASS magnitudes \citep[2\% and 5\%, respectively;][]{DR7paper,2mass}. The resulting values and fit are shown in Figures~\ref{fig:compchi} and \ref{fig:compimJ}. The corresponding equations are:
\begin{eqnarray*}
log_{10}(\chi) &=& (-0.0232\pm0.0022)\times(r'-K)^2 \\
&-& (0.0334\pm0.0442)\times(r'-K) - (3.8477\pm0.0292)\\
log_{10}(\chi) &=& (-0.0841\pm0.0091)\times(i-J)^2 \\
&-& (0.1301\pm0.0377)\times(i-J) - (3.7746\pm0.0343)
\end{eqnarray*}

We also fit quadratics to the SDSS M dwarf data for both colors, which yielded
\begin{eqnarray*}
log_{10}(\chi) &=& (-0.0226\pm0.0007)\times(r'-K)^2 \\
&-& (0.0374\pm0.0066)\times(r'-K) - (-3.8524\pm0.0148)\\
log_{10}(\chi) &=& (-0.0703\pm0.0033)\times(i-J)^2 \\
&-& (0.2025\pm0.0156)\times(i-J) - (3.6783\pm0.0175)
\end{eqnarray*}
Although the PHOENIX values are slightly below the SDSS values, the difference is $\lapprox0.05$ dex.  

\citet{schmidt2014} also calculate $\chi$ for $(i-J)>2.6$ (SpTs later than M7).  The $\chi$ values calculated here from SDSS and synthetic spectra are consistent with those from \citet{schmidt2014} in the overlapping range.

\begin{deluxetable*}{ccccccccccc}[!t]
\tablewidth{0pt}
\tabletypesize{\scriptsize}
\tablecaption{Colors and $\chi$ values from PHOENIX model spectra\label{phoenix}}
\tablehead{
\colhead{$T_{eff}$} & \colhead{SpT}  &  \colhead{$(g-r)$}  &  \colhead{$(r-i)$}  &  \colhead{$(i-z)$}  &  \colhead{$(z-J)$}  &  \colhead{$(J-H)$}  &  \colhead{$(H-K)$}  &  \colhead{$(r'-J)$}  &  \colhead{$(r'-K)$}  &  \colhead{$\chi$} \\
\colhead{(K)} & \colhead{}  &  \colhead{}  &  \colhead{}  &  \colhead{}  &  \colhead{}  &  \colhead{}  &  \colhead{}  &  \colhead{}  &  \colhead{}  &  \colhead{$(\times 10^{-5})$} 
}
\startdata
5200 & K0.3  &  0.660  &  0.192  &  0.069  &  0.935  &  0.491  &  0.025  &  1.183  &  1.699  &  9.170 \\ 
5100 & K0.9  &  0.704  &  0.207  &  0.082  &  0.954  &  0.513  &  0.029  &  1.229  &  1.771  &  9.633 \\ 
5000 & K1.4  &  0.751  &  0.223  &  0.094  &  0.974  &  0.536  &  0.033  &  1.277  &  1.847  &  9.592 \\ 
4900 & K2.0  &  0.800  &  0.241  &  0.108  &  0.995  &  0.562  &  0.038  &  1.328  &  1.928  &  9.421 \\ 
4800 & K2.6  &  0.852  &  0.260  &  0.122  &  1.018  &  0.591  &  0.043  &  1.384  &  2.018  &  9.292 \\ 
4700 & K3.3  &  0.911  &  0.281  &  0.137  &  1.042  &  0.619  &  0.050  &  1.443  &  2.112  &  9.209 \\ 
4600 & K3.9  &  0.974  &  0.305  &  0.154  &  1.065  &  0.645  &  0.057  &  1.507  &  2.210  &  9.142 \\ 
4500 & K4.4  &  1.037  &  0.333  &  0.174  &  1.086  &  0.668  &  0.066  &  1.575  &  2.309  &  9.087 \\ 
4400 & K4.8  &  1.096  &  0.366  &  0.196  &  1.104  &  0.684  &  0.077  &  1.647  &  2.408  &  8.673 \\ 
4300 & K5.3  &  1.153  &  0.403  &  0.222  &  1.122  &  0.695  &  0.089  &  1.726  &  2.510  &  8.211 \\ 
4200 & K6.0  &  1.205  &  0.442  &  0.250  &  1.138  &  0.698  &  0.105  &  1.807  &  2.610  &  7.645 \\ 
4100 & K6.7  &  1.249  &  0.484  &  0.278  &  1.155  &  0.696  &  0.122  &  1.894  &  2.711  &  7.156 \\ 
4000 & K7.9  &  1.287  &  0.533  &  0.312  &  1.174  &  0.689  &  0.140  &  1.994  &  2.823  &  6.836 \\ 
3900 & K9.3  &  1.318  &  0.589  &  0.351  &  1.197  &  0.682  &  0.156  &  2.110  &  2.948  &  6.533 \\ 
3800 & M0.3  &  1.343  &  0.654  &  0.393  &  1.224  &  0.677  &  0.170  &  2.241  &  3.088  &  6.211 \\ 
3700 & M0.9  &  1.361  &  0.732  &  0.442  &  1.255  &  0.672  &  0.180  &  2.397  &  3.248  &  5.867 \\ 
3600 & M1.5  &  1.376  &  0.820  &  0.494  &  1.290  &  0.667  &  0.187  &  2.568  &  3.422  &  5.502 \\ 
3500 & M2.1  &  1.391  &  0.916  &  0.548  &  1.329  &  0.662  &  0.194  &  2.754  &  3.610  &  5.026 \\ 
3400 & M2.7  &  1.408  &  1.021  &  0.604  &  1.375  &  0.657  &  0.200  &  2.957  &  3.814  &  4.483 \\ 
3300 & M3.3  &  1.430  &  1.134  &  0.663  &  1.427  &  0.653  &  0.208  &  3.177  &  4.039  &  3.918 \\ 
3200 & M3.9  &  1.453  &  1.264  &  0.727  &  1.489  &  0.651  &  0.218  &  3.430  &  4.298  &  3.331 \\ 
3100 & M4.5  &  1.474  &  1.419  &  0.799  &  1.562  &  0.652  &  0.227  &  3.724  &  4.603  &  2.716 \\ 
3000 & M5.1  &  1.485  &  1.603  &  0.882  &  1.643  &  0.656  &  0.235  &  4.065  &  4.957  &  2.183 \\ 
2900 & M5.6  &  1.494  &  1.817  &  0.974  &  1.746  &  0.659  &  0.243  &  4.466  &  5.368  &  1.703 \\ 
2800 & M6.3  &  1.489  &  2.068  &  1.078  &  1.866  &  0.663  &  0.248  &  4.933  &  5.844  &  1.252 \\ 
2700 & M7.2  &  1.480  &  2.346  &  1.189  &  1.999  &  0.668  &  0.252  &  5.445  &  6.365  &  0.886 \\ 
2600 & M8.0  &  1.566  &  2.539  &  1.294  &  2.142  &  0.675  &  0.257  &  5.879  &  6.811  &  0.618 \\ 
2500 & M8.5  &  1.757  &  2.623  &  1.371  &  2.269  &  0.682  &  0.257  &  6.164  &  7.103  &  0.472 
\enddata
\tablenotetext{*}{Spectral types are determined by interpolating the Teff-Spectral Type relationship assembled in table 5 of \citet{adam2007}.}
\tablecomments{Although we have given the colors as pairs of neighboring bands, better leverage on stellar properties is generally found with a wider spread in wavelength, e.g., $(i-J)$.  We computed synthetic photometry by convolving model spectra with the SDSS and 2MASS filter curves.}  
\end{deluxetable*}

\setlength{\baselineskip}{0.6\baselineskip}
\bibliography{references}
\setlength{\baselineskip}{1.667\baselineskip}

\end{document}